\documentclass[twocolumn,showpacs,secnumarabic,tightenlines,amssymb,
amsmath,nobibnotes,aps,prb]{revtex4}
\usepackage{graphicx}% Include figure files
\usepackage{dcolumn}% Align table columns on decimal point
\usepackage{bm}% bold math
\begin{document}
\preprint{APS/123-QED}
\title{New study of shot noise with the nonequilibrium Kubo formula 
in mesoscopic systems, \\
application to the Kondo effect at a quantum dot}
% Force line breaks with \\
\author{Tatsuya Fujii}
%\altaffiliation[Also at ]{}
%\author{aaa bbb$^{1,2}$}%
%\email{}
\affiliation{Institute for Solid State Physics, 
University of Tokyo, Kashiwa 277-8581, Japan}
%\author{}
%\homepage{}
%\affiliation{
\date{\today}
\begin{abstract}
Recently we have 
developed a theory of Keldysh formalism for 
mesoscopic systems. 
The resulting nonequilibrium Kubo formula 
for differential conductance makes it possible 
to propose the new formula of shot noise $S_h$, 
and thus to study shot noise in correlated systems 
at any temperature and any bias voltage. 
Employing this new approach, we analyze 
shot noise in the Kondo regime through a quantum dot 
for the symmetric case at zero temperature. 
Using the renormalized perturbation theory, 
we prove that in the leading order of bias voltage 
$S_h$ equal to noise power at zero temperature 
conventionally used as shot noise. 
With $S_h$, 
we calculate the Fano factor for a backscattering current 
$I_b$: $F_b=S_h/2eI_b$. 
It is shown that 
the Fano factor takes the universal form 
of $F_b=1+4(R-1)^2/(1+5(R-1)^2)$ 
determined by the Wilson ratio $R$ 
for arbitrary strength of the Coulomb interaction. 
Using the Wilson ratio $R=2$, 
our result coincides with the fractional value 
of $F_b=5/3$ already derived in the Kondo regime. 
\end{abstract}
%                              %display desired
\maketitle
\section{Introduction}

Mesoscopic devices have made it possible 
to investigate many-body effect in transport 
phenomena. 
A bias voltage applied to a conductor  
drives the system into a nonequilibrium steady state. 
These systems thus offer promising opportunities to 
study the correlated transport out of equilibrium. 

Shot noise has been introduced to explore 
nonequilibrium nature in these systems
\cite{beenakker, blanter, beenakker2}. 
Originally 
shot noise was given by the zero temperature value 
of noise power $S$ defined by the current-current 
correlation function. 
It is well known that, 
if uncorrelated particles with charge $q$ 
rarely transmit through a conductor, 
shot noise is given by the Poisson value 
$S=2qI$ where $I$ is the current. 
Therefore in this case, the quantity $S/2I$ has a possibility 
to determine the unit of charge carriers $q$, 
extracted from not only the current
\cite{beenakker, blanter, beenakker2}. 
In literature, this result is specific to 
a low transparent conductor without correlation effects. 
However, the possibility of shot noise 
has stimulated 
recent studies in other systems, for instance, 
correlated systems. 
There, the ratio $S/2I$ has been estimated as 
an indication of effective ''charge''. 
As a successful example to give 
a true quasiparticle charge,  
the fractional charge $e^*=e/3$ 
in the fractional quantum Hall regime
was determined through $S/2I$\cite{kane,samina}. 

Shot noise defined by the noise power has 
been useful to 
reveal attractive features 
at low temperatures. 
Extension of the concept of the shot noise 
at finite temperatures may 
open new possibilities out, 
however, how do we define shot noise 
at finite temperatures?

To illustrate the problem we take a quantum dot system 
as a stringent example\cite{Hershfield4}. 
In the noninteracting case, noise power splits
into thermal noise $S_t(T^0)$ and 
shot noise $S_s(T^0)$ as $S^0=S_t(T^0)+S_s(T^0)$ 
at any temperature
\cite{beenakker, blanter, beenakker2}, 
where $T^0$ is the bare transmission probability. 
The fact enables us to define the shot noise as 
$S^0-S_t(T^0)$ 
at any temperature and any bias voltage. 
With correlation effects, 
noise power is obtained as 
$S=S_t(T)+S_s(T)+\Delta S$. 
If the effect of Coulomb interaction 
may be included by simply 
changing $T^0$ to 
the full transmission probability $T$, 
noise power would be expressed as $S_t(T)+S_s(T)$. 
Shot noise thus could be defined in the same way 
as the uncorrelated system. 
However, an additional term of $\Delta S$ 
contributes to $S$. 
%%is given by the vertex-correction type diagram 
%%in the terminology of Feynman diagrams.  
From the expression of $\Delta S$, it is not 
possible to split it into 
the thermal-noise part and the shot-noise part. 
Thus in a quantum dot system it has not been 
possible to extract shot noise from the noise power. 
Until this point, we have stated 
the problem for a quantum dot system. 
However, 
the problem to define shot noise in correlated systems 
is a general one, not limited to the quantum dot 
system. 

Recently we have succeeded in extending 
a theory of Keldysh formalism for mesoscopic systems, 
and thus provided a basis for the definition of 
shot noise at any temperature. 

In this study, 
we have focused on a group of mesoscopic systems, 
typically designed as a correlated 
conductor attached to extended 
reservoirs through tunnel barriers. 
We have specifically examined 
the density matrix of the Keldysh formalism 
\cite{Keldysh,langreth,rammer}
for these systems 
\cite{Caroli,Hershfield,Meir}. 
An attempt to obtain an explicit expression 
of the density matrix had been already made in 
Refs.\onlinecite{Hershfield2,Hershfield3}. 
However, this attempt was not fully successful, since 
the density matrix still contained infinite series 
of operators which arose from the formal expansion 
of the s-matrix. 
Summing up them, we have thus proven that 
the density matrix has a form proposed by Maclennan 
and Zubarev
\cite{zubarevtxt,maclennan,TF071}. 
This type of density matrix was also derived 
by using a ${\rm C}^{*}$ algebra\cite{stasaki1,stasaki2}. 

The resulting density matrix allows us 
to derive a general expression of 
differential conductance: 
$G=\beta S/4 -\beta S_h/4$, 
where $S$ is the current-current correlation function, 
and $S_h$ is the non-trivial current-charge correlation 
function and $\beta=1/k_{\rm B}T$\cite{TF071}. 
We will call it 
the {\it nonequilibrium Kubo formula}. 
Furthermore, the density matrix has made it 
possible to prove that a steady state is realized, 
and describe 
the issue of irreversible processes in Keldysh formalism. 
Therefore our theory gives a unified description  
of desirable features 
in mesoscopic systems out of equilibrium. 

Moreover, the nonequilibrium Kubo formula 
allows us to introduce a natural definition 
of the shot noise 
in correlated systems. 
In fact the current-current 
correlation function $S$ is noise power, 
and of course $G$ is differential conductance. 
With these observable quantities, 
the nonequilibrium Kubo formula is  
written into $S_h=S-4k_{\rm B}TG$, and thus 
we propose the current-charge 
correlation function $S_h$ 
as the new definition of the shot noise. 
This concept has been confirmed in
a noninteracting quantum dot by explicitly calculating 
$S$, $G$ and the new formula $S_h$ with
the Keldysh Green function. 
The resulting relation among physical quantities 
is called as the {\it nonequilibrium identity}. 
Thus we can compare theoretically calculated 
$S_h$ at a certain temperature 
with $S-4k_{\rm B}TG$ 
determined from $S$ and $G$ measured in experiments.
Therefore, the nonequilibrium Kubo formula makes 
it possible to give 
a new perspective on studies of shot noise 
at any temperature in correlated systems. \cite{TF071}. 

Our aim in this paper is to 
apply this new approach to the quantum dot system 
in the Kondo regime by analyzing shot noise 
at low bias voltages. 

First, 
we briefly sketch recent studies of shot noise 
in the Kondo regime. 
In these studies shot noise is defined by 
the noise power $S$ at zero temperature. 
Theoretically 
shot noise was investigated 
in the s-d limit\cite{MeirGolub02}. 
When a bias voltage $eV$ is larger than the Kondo 
temperature $T_{\rm K}$, 
with decreasing the bias voltage  
shot noise increases logarithmically
as is well known in the Kondo systems. 
Close to the unitarity limit of $eV<T_{\rm K}$, 
shot noise is suppressed as 
%%%%%%%%%%%%%%%%%%%%%%%%%%%%%%%%%%%%%%%%%%%%%%%%%%%%%%%
\begin{eqnarray}
S = \frac{2e^2}{h}|V| 
\frac{\Gamma_L-\Gamma_R}{\Gamma_L+\Gamma_R}
+ \frac{4e^3 }{h}
  \frac{\gamma}{3}|V|
\left( \frac{eV}{T_{\rm K}} \right)^2 +\cdots, 
\label{in1}
\end{eqnarray}
%%%%%%%%%%%%%%%%%%%%%%%%%%%%%%%%%%%%%%%%%%%%%%%%%%%%%%%
where $\Gamma_{L,R}$ are the resonance widths 
between dot and left and right leads. 
The Fermi-liquid fixed point Hamiltonian $H$ 
\cite{Nozieres78Hf,Glatxmantxt04} is used 
in calculations. 

In recent studies\cite{Sela06,Golub06}, 
the Fermi-liquid features in eq.(\ref{in1}) 
have been reexamined at zero temperature. 
The fixed point Hamiltonian is 
characterized by a one-particle scattering  
and a two-particle scattering. 
They give corrections to the $\pi/2$ phase shift 
in the unitarity limit: 
a frequency-liner phase shift and 
a quasiparticle-distribution-linear phase shift. 

As a result, a current 
$I_b \equiv -e \partial_t (N_{LH}(t)-N_{RH}(t))$
describes a reduction from the perfect 
transmission for the $\pi/2$ phase shift 
in the unitarity limit, 
which is the definition of 
the backscattering current in 
Refs.\onlinecite{Sela06,Golub06}. 
Here $N_{L,R}$ express particle-number operators 
in left and right leads and $A_{H}(t)$ is 
the Heisenberg representation for an operator $A$. 

At zero temperature, the observable current 
is defined by $I=2e^2 V/h -I_b$, 
where $I_u =2e^2 V/h$ is 
the current in the unitarity limit. 
Since $2e^2 V/h$ becomes a constant value, 
noise power given by 
the current-current correlation function 
is rewritten as 
%%%%%%%%%%%%%%%%%%%%%%%%%%%%%%%%%%%%%%%%%%%%%%%%%%%%%%%
\begin{eqnarray}
S=\int^{\infty}_{-\infty}dt 
\langle \{ \delta I_{bH}(t),\delta I_{bH}(0)\} \rangle.
\label{in2}
\end{eqnarray}
%%%%%%%%%%%%%%%%%%%%%%%%%%%%%%%%%%%%%%%%%%%%%%%%%%%%%%%
Therefore the shot noise is determined by only 
the backscattering current when $eV<T_{\rm K}$. 

On the basis of this fact, using $S/2I_b$, 
it has been attempted to estimate an effective 
''charge'' for the backscattering current\cite{Sela06}. 
In the present work we will discuss 
the Fano factor $F_b=S/2eI_b$ 
in units of charge $e$

In Ref.\onlinecite{Sela06},  
the symmetric case 
$\Gamma_L=\Gamma_R$ is especially focused on. 
This condition makes the leading order of $S$ 
in the bias voltage 
${\cal{O}}(V^3)$ as in eq.(\ref{in1}) . 
Then, the leading order of $I_b$ also becomes 
${\cal{O}}(V^3)$. 
In the symmetric case, 
$F_b=S/2eI_b$ results in a 
universal fractional value up to ${\cal{O}}(V^3)$
%%%%%%%%%%%%%%%%%%%%%%%%%%%%%%%%%%%%%%%%%%%%%%%%%%%%%%%
\begin{eqnarray}
F_b =\frac{S}{2eI_b}=\frac{5}{3}, 
\label{in3}
\end{eqnarray}
%%%%%%%%%%%%%%%%%%%%%%%%%%%%%%%%%%%%%%%%%%%%%%%%%%%%%%%T
It has been pointed out that this universal 
enhancement from the unity 
originates from the two-particle backscattering. 

This universal feature has stimulated 
further studies: in the context of 
the full counting statistics\cite{Gogolin06} 
this result has been reproduced, 
and a shot-noise 
measurement has been reported on $F_b =5/3$ 
close to the unitarity limit at low bias 
voltages\cite{Zarchin08}. We address this topic 
from our point of view. 

In this paper, we begin with a brief review of 
the generalization of the Kubo formula into 
a nonequilibrium situation and shot noise 
in Sec.{\ref{review}}. 
In Sec.{\ref{shkqd}} applying our new concept 
to a quantum dot system 
described by the Anderson model, 
we investigate the shot noise and 
the Fano factor in the Kondo regime. 
We concentrate on 
the symmetric case at zero temperature. 
At zero temperature the nonequilibrium 
Kubo formula yields $S=S_h$. 
Thus, using the renormalized perturbation 
theory, we calculate noise power $S$ and 
shot noise $S_h$ independently 
up to the leading order for the symmetric 
case: ${\cal{O}}(V^3)$ and show that 
they are identical. 
The backscattering current is also calculated 
up to ${\cal{O}}(V^3)$.
Finally 
we will discuss the universal feature of 
the Fano factor 
for any Coulomb interaction 
up to ${\cal{O}}(V^3)$. 
\section{shot noise and nonequilibrium Kubo formula}\label{review}

We start with a brief summary of shot noise and 
the nonequilibrium Kubo formula based on Keldysh formalism 
for mesoscopic systems in nonequilibrium 
steady states\cite{TF071}.

\subsection{density matrix}\label{density}
Our starting point is the Hamiltonian for 
a class of mesoscopic systems: 
a correlated conductor $H_{c}$ attached to left and 
right infinite reservoirs $H_{L,R}$ through 
boundary couplings $H_{cL,R}$. 
For instance a quantum dot system 
is a typical example of them. 
The Hamiltonian is described as 
%%%%%%%%%%%%%%%%%%%%%%%%%%%%%%%%%%%%%%%%%%%%%%%%%%%%%%%
\begin{eqnarray}
H \equiv H_{c} + H_{L} + H_{R} + H_{cL,R}. 
\label{hamiltonian0}
\end{eqnarray}
%%%%%%%%%%%%%%%%%%%%%%%%%%%%%%%%%%%%%%%%%%%%%%%%%%%%%%%%
$H_{c}$ is given by 
noninteracting and interacting terms 
as $H_c \equiv H_{c0}+H_{c1}$. 
Each reservoir must be sufficiently large 
to behave as a good thermal bath. 
%Within this condition, 
%any $H_{L,R}$ would be possible. 

In a conventional explanation of Keldysh formalism, 
it seems that concepts such as 
the Keldysh contour, and the Keldysh Green 
function are stressed\cite{Keldysh}. 
In contrast, our main argument is to examine 
the early stage of Keldysh formalism where 
the perturbation term is adiabatically 
turned on. 

To begin with,  we divide $H$ into 
an unperturbed term $H_0 \equiv H_{c0}+H_{L}+H_{R}$ 
and a perturbation term $H_1 \equiv H_{cL,R}+H_{c1}$ 
\cite{Caroli,Hershfield,Meir}. 
To adiabatically introduce the perturbation term, 
we use a time-dependent Hamiltonian 
as $H_{{\epsilon} t} \equiv H_0 + e^{-\epsilon |t|}H_1$. 
%%
%%Keldysh formalism implicitly assumes that 
%%infinite perturbation series of the s-matrix 
%%is assumed to converge. 
%%it is worth to remind that $H_0$ describes 
%%the infinite system, while $H_1$ does the finite system. 
%%Thus it might be impossible for the perturbation of the finite system 
%%to force the infinite system to occur a phase transition. 
%%even if a finite system perturbs an infinite system, 
%%a phase transition is not expected to occur. 
%%it means that infinite series of the s-matrix converges. 
%%It is rather worth to point out that 
%%the energy scale of $H_0$ for the infinite system 
%%is much larger than one of $H_1$ for the finite system. 
%%The energy scale of $H_0$ 
%%is expected to be much larger than one of $H_1$. 
%%

The initial state at $t_0=-\infty$ is 
determined by the three separated systems in $H_0$: 
left and right reservoirs and the conductor 
which have different chemical 
potentials $\mu_{L,R}$, $\mu_c$. 
We define 
a bias voltage as $eV/2 \equiv (\mu_L-\mu_R)/2$ 
and an averaged chemical potential 
$\mu \equiv (\mu_L+\mu_R)/2$, 
where $\mu = \mu_c$ is assumed. 
The origin of energy is set to $\mu = \mu_c=0$. 
As a result the initial density matrix is given by 
%%%%%%%%%%%%%%%%%%%%%%%%%%%%%%%%%%%%%%%%%%%%%%%%%%%%%%%
\begin{eqnarray}
\rho _{0} = e^{-\beta (H_0 -eV/2 (N_L -N_R) -\Omega _0)}. 
\label{inidens}
\end{eqnarray}
%%%%%%%%%%%%%%%%%%%%%%%%%%%%%%%%%%%%%%%%%%%%%%%%%%%%%%%%
$[H_0, N_L -N_R ]=0$ is assumed throughout. 
The initial thermodynamic potential is chosen 
to satisfy ${\rm Tr}\{\rho _{0}\}=1$, 
$\Omega _0 =-1/\beta  \ln {\rm Tr} 
e^{-\beta (H_0 -eV/2 (N_L -N_R)) }$. 

The time evolution of the density matrix $\rho _{\epsilon} (t)$ 
is determined by the Neumann equation. 
%%%%%%%%%%%%%%%%%%%%%%%%%%%%%%%%%%%%%%%%%%%%%%%%%%%%%%%
%%\begin{eqnarray}
%%\rho _{\epsilon}(t)
%%= e^{-{\rm i}H_0 t} S_{\epsilon}(t,t_0) e^{{\rm i}H_0 t_0}
%%\rho _{\epsilon}(t_0) 
%%e^{-{\rm i}H_0 t_0} S_{\epsilon}(t_0,t) e^{{\rm i}H_0 t}. 
%%\label{formalsol}
%%\end{eqnarray}
%%%%%%%%%%%%%%%%%%%%%%%%%%%%%%%%%%%%%%%%%%%%%%%%%%%%%%%%
With $\rho _{\epsilon} (t)$, 
the expectation value of any operator $\cal O$ is defined by 
${\rm Tr} \{ \rho _{\epsilon} (t) {\cal O} \}$. 
It is transformed as 
%%%%%%%%%%%%%%%%%%%%%%%%%%%%%%%%%%%%%%%%%%%%%%%%%%%%%%%
\begin{eqnarray}
{\rm Tr} \{ \rho _{\epsilon} (t) {\cal O} \}
= {\rm Tr} \{ {\rho}_{\epsilon} (0) {\cal O}_H (t) \},
\label{exv}
\end{eqnarray}
%%%%%%%%%%%%%%%%%%%%%%%%%%%%%%%%%%%%%%%%%%%%%%%%%%%%%%%%
where the Heisenberg representation is given by 
${\cal O}_H (t)=S_{\epsilon}(0,t) {\cal O}(t) S_{\epsilon}(t,0)$
and ${\cal O}(t)=e^{{\rm i}H_0 t} {\cal O}  e^{-{\rm i}H_0 t}$. 

Here in order to avoid confusion, 
we comment on 
the time dependence of the expectation value. 
It is determined by only the Heisenberg representation 
of an operator ${\cal O}$ as shown 
in the right-hand side of eq.(\ref{exv}). 
Concerning the density matrix, 
it is sufficient to consider ${\rho}_{\epsilon} (0)$ at $t=0$. 
We thus consider a specific time of $t=0$, 
but until this point we do not use any assumption 
either the steady state is realized at $t=0$ or not. 
After taking a limit of $\epsilon \rightarrow 0$, 
the system reaches a steady state. 
Later we will return back to this point 
in Sec.{\ref{invariant}}. 

We thus focus on the density matrix 
${\rho}_{\epsilon} (0) \equiv \bar{\rho}_{\epsilon}$ 
which is given by a formal solution of the 
Neumann equation 
%%%%%%%%%%%%%%%%%%%%%%%%%%%%%%%%%%%%%%%%%%%%%%%%%%%%%%%
\begin{eqnarray}
\bar{\rho}_{\epsilon}
= S_{\epsilon}(0,-\infty) \rho _{0} S_{\epsilon}(-\infty ,0). 
\label{rho01}
\end{eqnarray}
%%%%%%%%%%%%%%%%%%%%%%%%%%%%%%%%%%%%%%%%%%%%%%%%%%%%%%%%
Using the unitarity of 
$S_{\epsilon}(0,-\infty)S_{\epsilon}(-\infty ,0)=1$, 
$\bar{\rho}_{\epsilon}$ becomes 
%%%%%%%%%%%%%%%%%%%%%%%%%%%%%%%%%%%%%%%%%%%%%%%%%%%%%%%
\begin{eqnarray}
\nonumber
&&{\!\!\!\!\!\!\!\!\!\!\!}
\bar{\rho}_{\epsilon}
= {\rm exp} \{ -\beta ( 
S_{\epsilon}(0,-\infty) H_0 S_{\epsilon}(-\infty ,0) \\
&& 
-V S_{\epsilon}(0,-\infty) e/2 (N_L -N_R) S_{\epsilon}(-\infty ,0)
-\Omega _0)  \}. 
\label{formal}
\end{eqnarray}
%%%%%%%%%%%%%%%%%%%%%%%%%%%%%%%%%%%%%%%%%%%%%%%%%%%%%%%%
To proceed calculations, 
we have to compute a type of quantity
%%%%%%%%%%%%%%%%%%%%%%%%%%%%%%%%%%%%%%%%%%%%%%%%%%%%%%%
\begin{eqnarray}
\bar{A}_{\epsilon} \equiv
S_{\epsilon}(0,-\infty) A 
S_{\epsilon}(-\infty,0). 
\label{abar}
\end{eqnarray}
%%%%%%%%%%%%%%%%%%%%%%%%%%%%%%%%%%%%%%%%%%%%%%%%%%%%%%%%
We have two cases in our mind: $A=H_0$ and $A=e/2(N_L-N_R)$, 
where 
we suppose that $A$ satisfies $[A,H_0]=0$. 
The critical step in calculations is to derive
\cite{TF071}
%%%%%%%%%%%%%%%%%%%%%%%%%%%%%%%%%%%%%%%%%%%%%%%%%%%%%%%
\begin{eqnarray}
\bar{A}_{\epsilon}
=A+ \int _{-\infty}^{0} {\rm d}t e^{-\epsilon |t|} J_{AH}(t). 
\label{adiabatic}
\end{eqnarray}
%%%%%%%%%%%%%%%%%%%%%%%%%%%%%%%%%%%%%%%%%%%%%%%%%%%%%%%%
The "current" for the operator $A$ is defined as 
%%%%%%%%%%%%%%%%%%%%%%%%%%%%%%%%%%%%%%%%%%%%%%%%%%%%%%%
\begin{eqnarray}
J_{AH}(t) \equiv -\frac{\partial}{\partial t} A_H (t).
\label{fcurrent}
\end{eqnarray}
%%%%%%%%%%%%%%%%%%%%%%%%%%%%%%%%%%%%%%%%%%%%%%%%%%%%%%%%

Technically eq.(\ref{adiabatic}) is sufficient 
to calculate the density matrix. 
However, we shortly discuss eq.(\ref{adiabatic}), 
leading to deeper understanding of
the nonequilibrium nature in Keldysh formalism.
Integration by parts in eq.(\ref{adiabatic}) yields 
%%%%%%%%%%%%%%%%%%%%%%%%%%%%%%%%%%%%%%%%%%%%%%%%%%%%%%%
\begin{eqnarray}
{\quad}
\bar{A}_{\epsilon}
= \epsilon \int _{-\infty}^{0} {\rm d}t e^{-\epsilon |t|} A_{H}(t). 
\label{adiabatic2}
\end{eqnarray}
%%%%%%%%%%%%%%%%%%%%%%%%%%%%%%%%%%%%%%%%%%%%%%%%%%%%%%%%
Thus $\bar{A}_{\epsilon}$ expresses a long-time 
average of $A$ in the limit 
$\epsilon \rightarrow 0$. 
In fact, $\bar{A}_{\epsilon}$ defined by eq.(\ref{adiabatic2}) 
represents nothing but the invariant part of 
an operator $A$, 
introduced by Zubarev\cite{zubarevtxt}. 
We have thus proven that the adiabatic 
introduction of $H_1$ in Keldysh formalism 
corresponds to taking 
the invariant part by Zubarev. 
We revisit this concept to discuss steady states 
and irreversible processes in Sec.{\ref{invariant}}. 

We use eq.(\ref{adiabatic}) 
in two cases: $A=H_0$ and $A=e/2(N_L -N_R)$ where 
the "currents" are defined 
by energy change $J_{eH}(t)$ and charge current $J_{cH}(t)$ 
using eq.(\ref{fcurrent}), respectively. 
Substituting them into eq.(\ref{formal}) yields 
%%%%%%%%%%%%%%%%%%%%%%%%%%%%%%%%%%%%%%%%%%%%%%%%%%%%%%%
\begin{eqnarray}
\nonumber
&&{\!\!\!\!\!\!\!\!\!}
\bar{\rho}_{\epsilon} 
={\rm exp} \left\{ -\beta(H_0 + \int ^{0}_{-\infty}
{\rm d}t e^{-\epsilon |t|}  J_{eH}(t) \right. \\
&&{\!\!\!}- \left. eV/2(N_L -N_R) -V \int ^{0}_{-\infty}
{\rm d}t e^{-\epsilon |t|}  J_{cH}(t) -\Omega _0) \right \}.
\nonumber \\
\label{zubarev}
\end{eqnarray}
%%%%%%%%%%%%%%%%%%%%%%%%%%%%%%%%%%%%%%%%%%%%%%%%%%%%%%%%
The density matrix of 
Keldysh formalism thus becomes a type of 
the nonequilibrium statistical operators 
initiated by MacLennan and Zubarev\cite{maclennan,zubarevtxt}. 
This form was also obtained by 
the ${\rm C}^{*}$ algebra\cite{stasaki1,stasaki2}. 
\subsection{nonequilibrium Kubo formula}\label{nkf}

The resultant density matrix makes it possible 
to generalize the Kubo formula for 
conductance into a nonequilibrium 
steady state. 
For convenience, in this work we introduce 
a bra-ket notation: 
$\langle O \rangle \equiv 
\displaystyle \lim _{\epsilon \rightarrow 0} 
{\rm Tr}\{ \bar{\rho}_{\epsilon} {\cal O} \}$. 
The current is thus denoted as 
%%%%%%%%%%%%%%%%%%%%%%%%%%%%%%%%%%%%%%%%%%%%%%%%%%%%%%%
\begin{eqnarray}
I \equiv 
\langle J_{H}(t) \rangle = 
\langle J \rangle, 
\label{recurrent}
\end{eqnarray}
%%%%%%%%%%%%%%%%%%%%%%%%%%%%%%%%%%%%%%%%%%%%%%%%%%%%%%%%
where in the last equality of eq.(\ref{recurrent}), 
a steady-state feature later proved 
in Sec.{\ref{invariant}} is used. 
From the current, differential conductance is given as 
%%%%%%%%%%%%%%%%%%%%%%%%%%%%%%%%%%%%%%%%%%%%%%%%%%%%%%%
\begin{eqnarray}
G \equiv \frac{\partial I}{\partial V}. 
\label{conductance}
\end{eqnarray}
%%%%%%%%%%%%%%%%%%%%%%%%%%%%%%%%%%%%%%%%%%%%%%%%%%%%%%%%
Eq.(\ref{recurrent}) shows that 
the current $I$ depend on a bias voltage $V$ 
only through the density matrix $\bar{\rho}_{\epsilon}$. 
To obtain $G$, 
$\partial \bar{\rho}_{\epsilon} /\partial V$ should be
calculated. 
%%$H_0$ implicitly includes $V$-dependent 
%%potential terms in reservoirs. 
%%Explicitly, there is the $V$-linear term 
%%in the second line of eq.(\ref{zubarev}). 
%%Moreover, the initial thermodynamic potential 
%%$\Omega _0$ is also dependent $V$. 
%%In a wide-band limit $\partial H_0/\partial V$ 
%%becomes sufficiently legible. 
%%Thus relevant contributions 
%%come from the $V$-linear term 
%%and $\Omega _0$
%%in the second line of eq.(\ref{zubarev}). 
Thus $G$ is given by 
%%%%%%%%%%%%%%%%%%%%%%%%%%%%%%%%%%%%%%%%%%%%%%%%%%%%%%%
\begin{eqnarray}
\nonumber
&&G=\lim _{\epsilon \rightarrow 0}
\beta {\rm Tr}  \biggl\{  
J {\bigg(} e/2(N_L -N_R)  \\
&&{\qquad \qquad } \left. \left. 
+ \int^{0}_{-\infty} {\!\!\!} {\rm d}t e^{-\epsilon |t|} 
J_{H}(t) + \frac{\partial \Omega _0}{\partial V}
\right) \bar{\rho}_{\epsilon}  \right\}. 
\label{gv1}
\end{eqnarray}
%%%%%%%%%%%%%%%%%%%%%%%%%%%%%%%%%%%%%%%%%%%%%%%%%%%%%%%%
Using the initial thermodynamic 
potential $\Omega _0$ and 
$S_{\epsilon}S^{\dagger}_{\epsilon}=1$, 
$\partial \Omega _0 /\partial V$ is calculated as 
%%%%%%%%%%%%%%%%%%%%%%%%%%%%%%%%%%%%%%%%%%%%%%%%%%%%%%%
\begin{eqnarray}
\nonumber
\lim _{\epsilon \rightarrow 0}
\frac{\partial \Omega _0}{\partial V} = 
-e/2\langle N_L -N_R \rangle   
+ \int^{0}_{-\infty} {\!\!\!} {\rm d}t e^{-\epsilon |t|} 
\langle J_{H}(t) \rangle .
\label{difomega0}
\end{eqnarray}
%%%%%%%%%%%%%%%%%%%%%%%%%%%%%%%%%%%%%%%%%%%%%%%%%%%%%%%%
Thus the inside of $(...)$ in eq.(\ref{gv1}) 
is expressed by a fluctuation of an operator: 
$\delta \bar{A}_{\epsilon} 
\equiv \bar{A}_{\epsilon} - 
\langle \bar{A}_{\epsilon} \rangle$. 
In fact, $\bar{A}_{\epsilon}$ just expresses
the invariant part of $A=e/2(N_L-N_R)$ 
in eq.(\ref{adiabatic}). 
For simplicity 
we abbreviate it $\bar{A}_{\epsilon}$ for a while. 
Furthermore, using 
$\langle B \delta C \rangle =
\langle \delta B \delta C \rangle$ 
for any operators $B$ and $C$, 
eq.(\ref{gv1}) is rewritten into 
$
G=\beta \langle \delta J \delta \bar{A}_{\epsilon} \rangle
$

Here we symmetrize the expression of 
$
G=\beta \langle \delta J \delta \bar{A}_{\epsilon} \rangle
$. 
The current operator $J$ and 
$\bar{A}_{\epsilon}$ do not commute, 
but it is possible to transform to 
the inverse order of 
$\langle \delta \bar{A}_{\epsilon} \delta J \rangle$. 
Using 
$S_{\epsilon}[H_0,e/2(N_L-N_R)]S^{\dagger}_{\epsilon}=0$, 
$\bar{A}_{\epsilon}$ 
is proven to commute with $\bar{\rho}_{\epsilon}$. 
With a cyclic permutation inside the trace, 
$
G=\beta \langle \delta J \delta \bar{A}_{\epsilon} \rangle
$
can be written 
into 
$
G=\beta \langle \delta \bar{A}_{\epsilon} \delta J  \rangle
$. 
We have thus obtain two different expression of $G$. 

Moreover each form of $G$ is characterized by 
the time-integral over negative $t$. 
It is rewritten into one over positive $t$ using 
the steady-state feature which will be discussed 
in eq.(\ref{st}). 
Consequently, 
four different expressions of $G$ are derived. 
Symmetrizing them yields 
%%%%%%%%%%%%%%%%%%%%%%%%%%%%%%%%%%%%%%%%%%%%%%%%%%%%%%%
\begin{eqnarray}
\nonumber
&&G=
\frac{\beta}{4}
\langle \{ \delta J , e(\delta N_L -\delta N_R) \}\rangle  \\
&&{\qquad \qquad }
+
\frac{\beta}{4}
\int^{\infty}_{-\infty} {\!\!\!} {\rm d}t 
%%e^{-\epsilon |t|} 
\langle \{ \delta J_{H}(t), \delta J_{H}(0) \} \rangle ,
% \delta J_{cH}(0) \delta J_{cH}(t) \rangle , 
\label{symgv}
\end{eqnarray}
%%%%%%%%%%%%%%%%%%%%%%%%%%%%%%%%%%%%%%%%%%%%%%%%%%%%%%%%
where $\{B,C\}=BC+CB$ represents the anticommutation relation. 

We have found that differential conductance $G$ is 
determined by 
a current-current correlation function, 
and an unusual current-charge correlation function. 
In the linear response regime for $G$: $V=0$, 
we can show that 
%%%%%%%%%%%%%%%%%%%%%%%%%%%%%%%%%%%%%%%%%%%%%%%%%%%%%%%%
\begin{eqnarray}
\langle \{ \delta J , 
e(\delta N_L -\delta N_R) \}\rangle_{V=0} =0, 
\end{eqnarray}
%%%%%%%%%%%%%%%%%%%%%%%%%%%%%%%%%%%%%%%%%%%%%%%%%%%%%%%%
and thus eq.(\ref{symgv}) reduces to the standard 
Kubo formula. 
Away from the linear response regime, 
the current-current correlation function 
gives a naive extension 
from the relation between fluctuation and dissipation. 
On the other hand, 
the current-charge correlation function 
gives an intriguing modification 
far from the linear response regime. 
In conclusion we have succeeded in obtaining a 
{\it nonequilibrium Kubo formula} in mesoscopic systems. 
\subsection{shot noise}\label{shot}

We reexamine the nonequilibrium Kubo formula 
in view of physical quantities. 
As measurable quantities, 
there are differential conductance $G$ 
and noise power given by 
the current-current correlation function $S$
%%%%%%%%%%%%%%%%%%%%%%%%%%%%%%%%%%%%%%%%%%%%%%%%%%%%%%%
\begin{eqnarray}
S=\int^{\infty}_{-\infty}dt 
\langle \{ \delta J_H(t),\delta J_H(0)\} \rangle .
\label{npdf}
\end{eqnarray}
%%%%%%%%%%%%%%%%%%%%%%%%%%%%%%%%%%%%%%%%%%%%%%%%%%%%%%%%
At a steady state noise power at zero frequency 
dominates. 

On the other hand, we define the current-charge 
correlation function as $S_h$ 
%%%%%%%%%%%%%%%%%%%%%%%%%%%%%%%%%%%%%%%%%%%%%%%%%%%%%%%
\begin{eqnarray}
S_h= 
-\langle \{ \delta J,e(\delta N_L -\delta N_R) \} \rangle .
\label{shdf}
\end{eqnarray}
%%%%%%%%%%%%%%%%%%%%%%%%%%%%%%%%%%%%%%%%%%%%%%%%%%%%%%%%

The nonequilibrium Kubo  formula in eq.(\ref{symgv}) 
leads to the relation among $S_h$ and 
two observable quantities $G$ and $S$
%%%%%%%%%%%%%%%%%%%%%%%%%%%%%%%%%%%%%%%%%%%%%%%%%%%%%%%
\begin{eqnarray}
S_h = S-4 k_{\rm B} T G.
\label{shsg}
\end{eqnarray}
%%%%%%%%%%%%%%%%%%%%%%%%%%%%%%%%%%%%%%%%%%%%%%%%%%%%%%%%

As stated in the introduction, 
in uncorrelated systems shot noise 
is given by $S^0-S_t(T^0)$. 
Considering the fact that the thermal noise 
$S_t(T^0)=4 k_{\rm B} T G^0$, 
$S_h$ in eq.(\ref{shsg}) suggests 
the natural extension of the shot noise. 
Therefore, we propose to define the shot noise by 
the current-charge correlation function $S_h$
at arbitrary temperature and bias voltage 
in correlated systems. 

As a result, eq.(\ref{shsg}) expresses 
the relation among physical quantities 
out of equilibrium. Thus we will call it 
a {\it nonequilibrium identity}. 
In the linear response regime, 
as discussed previously 
the current-charge correlation function vanishes, 
and thus $S_h =0$. 
Eq.(\ref{shsg}) then goes back to 
the Nyquist-Johnson relation $S(0)=4 k_{\rm B} T G(0)$.
On the other hand, at zero temperature 
eq.(\ref{shsg}) shows that 
$S_h$ indeed equals to $S$ at $T=0$ 
which is nothing but the original definition of 
the shot noise. 

\subsection{steady state}\label{invariant}

Having discussed consequence of 
the density matrix on the transport property: 
the nonequilibrium Kubo formula and shot noise, 
here we turn to others: 
irreversible processes and steady states 
in Keldysh formalism. 

We have shown that eq.(\ref{adiabatic}), 
more explicitly eq.(\ref{adiabatic2}) 
corresponds to the invariant part of 
an operator $A$ discussed by Zubarev.
Originally, this concept was introduced 
in a local equilibrium system. 
Zubarev could show that 
the entropy production was positive 
in the linear response regime, and thus concluded 
that the method could describe dissipations. 
Following the same analysis, 
it is possible to prove that 
Keldysh formalism have the same feature 
as Zubarev theory. 

Secondly we discuss steady states. 
We have shown in Ref.\onlinecite{TF071} that 
the commutation relation between 
the invariant part and the Hamiltonian satisfies 
%%%%%%%%%%%%%%%%%%%%%%%%%%%%%%%%%%%%%%%%%%%%%%%%%%%%%%%
\begin{eqnarray}
\lim_{\epsilon \rightarrow 0}
[\bar{A}_{\epsilon}, H_{\epsilon t}] 
= 
\lim_{\epsilon \rightarrow 0}
-{\rm i} \epsilon (\bar{A}_{\epsilon}-A). 
\label{commutation7}
\end{eqnarray}
%%%%%%%%%%%%%%%%%%%%%%%%%%%%%%%%%%%%%%%%%%%%%%%%%%%%%%%%
Assuming that $\bar{A}_{\epsilon}$ exists 
in the limit $\epsilon \rightarrow 0$, 
the r.h.s. of eq.(\ref{commutation7}) vanishes. 
In $\epsilon \rightarrow 0$ 
the invariant part commutes with the Hamiltonian, 
and thus conserves. 

This result is crucial for the proof 
that a steady state is realized in the limit 
$\epsilon \rightarrow 0$. 
Using a similar technique to derive 
eq.(\ref{adiabatic}), 
the expectation value of an operator $\cal O$ 
is rewritten into 
%%%%%%%%%%%%%%%%%%%%%%%%%%%%%%%%%%%%%%%%%%%%%%%%%%%%%%%
\begin{eqnarray}
\nonumber
&&\lim_{\epsilon \rightarrow 0}
{\rm Tr}\{ \bar{\rho}_{\epsilon} {\cal O}_H(t) \} 
=
\lim_{\epsilon \rightarrow 0}
{\rm Tr}\{ \bar{\rho}_{\epsilon} {\cal O} \}\\
&&{\quad}+ \lim_{\epsilon \rightarrow 0}
\int^{t}_{0} {\rm d}t^{'} {\rm Tr}\{ 
{\rm i}[\bar{\rho}_{\epsilon}, H_{\epsilon t^{'}}]
J_{{\cal O}H} (t-t^{'}) \}. 
%%
%%S_{\epsilon}(0,t_1) {\rm e}^{{\rm i} H_0 t_1}
%%%{\rm i}[\bar{\rho}_{\epsilon}, H_{\epsilon 1}] 
%%{\rm e}^{-{\rm i} H_0 t_1} S_{\epsilon}(t_1,0)
%%J_{{\cal O} H}(t) \}. 
\label{exp1}
\end{eqnarray}
%%%%%%%%%%%%%%%%%%%%%%%%%%%%%%%%%%%%%%%%%%%%%%%%%%%%%%%%
The definition of $\bar{\rho}_{\epsilon}$ in 
eq.(\ref{rho01}) shows that 
the density matrix itself in fact becomes an invariant part. 
As a consequence the commutation relation 
$[\bar{\rho}_{\epsilon}, H_{\epsilon t^{'}}]$ 
in eq.(\ref{exp1}) vanishes 
in the limit $\epsilon \rightarrow 0$, leading to 
%%%%%%%%%%%%%%%%%%%%%%%%%%%%%%%%%%%%%%%%%%%%%%%%%%%%%%%
\begin{eqnarray}
\lim_{\epsilon \rightarrow 0}
{\rm Tr}\{ \bar{\rho}_{\epsilon} {\cal O}_H(t) \}
=\lim_{\epsilon \rightarrow 0} 
{\rm Tr}\{ \bar{\rho}_{\epsilon} {\cal O} \}. 
\label{st}
\end{eqnarray}
%%%%%%%%%%%%%%%%%%%%%%%%%%%%%%%%%%%%%%%%%%%%%%%%%%%%%%%%
We have proven that 
a steady state is realized in the limit $\epsilon \rightarrow 0$. 
According to the same analysis, 
$\langle {\cal O}_H(t){\cal O}_H(t') \rangle =
\langle {\cal O}_H(t-t'){\cal O}_H(0) \rangle $ 
can be also derived. 
\section{shot noise in the Kondo 
effect at a quantum dot}\label{shkqd}
\subsection{Quantum dot system and physical quantities}

Now let us turn to the discussion on 
shot noise through a quantum dot in the Kondo regime. 
We consider the Anderson model with a single-level, 
%%%%%%%%%%%%%%%%%%%%%%%%%%%%%%%%%%%%%%%%%%%%%%%%%%%%%%%
\begin{eqnarray}
 H= \sum_{k\alpha\sigma} 
     {\varepsilon}_{k \alpha} 
      c^{\dagger}_{k\alpha\sigma} c_{k\alpha\sigma}
      +\sum_{\sigma} {\epsilon}_d n_{\sigma}
       +U n_{\uparrow} n_{\downarrow} & &
             \nonumber\\
      + \sum_{k\alpha\sigma}
          \left(V_{k\alpha} c^{\dagger}_{k\alpha\sigma} d_{\sigma}
           +{\rm h.c.}\right).& &
\label{dot}
\end{eqnarray}
%%%%%%%%%%%%%%%%%%%%%%%%%%%%%%%%%%%%%%%%%%%%%%%%%%%%%%%%
$d^{\dagger}_{\sigma}$ and 
$c^{\dagger}_{k\alpha\sigma}(\alpha=L,R)$ 
create an electron 
with spin $\sigma$ at the dot and the left-right lead respectively. 
The last term describes the tunneling between the dot 
and the leads, which determines the resonance width 
$\Gamma (\omega)=
(\Gamma_{L}(\omega)+\Gamma_{R}(\omega))/2$
with 
$\Gamma_{L,R}(\omega)=2\pi\sum_k |V_{kL,R}|^2 \delta 
(\omega -\varepsilon_{k L,R})$. 
In the limit of large band width, 
the resonance width may be assumed to be a constant 
$\Gamma=(\Gamma_{L}+\Gamma_{R})/2$. 

As discussed in the introduction, 
we focus on the Kondo effect 
near the unitarity limit at zero temperature, 
and investigate shot noise up to ${\cal{O}}(V^3)$. 
To discuss the unitarity limit, 
it is natural to concentrate on the symmetric case,
${\epsilon}_d=-U/2$ and $\Gamma_{L}=\Gamma_{R}=\Gamma$. 

Traditionally noise power $S$ at zero temperature 
has been defined as shot noise. In contrast 
we define the shot noise in general by eq.(\ref{shdf}). 
From the nonequilibrium identity 
eq.(\ref{shsg}) based on the nonequilibrium Kubo formula, 
%
%%%%%%%%%%%%%%%%%%%%%%%%%%%%%%%%%%%%%%%%%%%%%%%%%%%%%%%
\begin{eqnarray}
S_h=S, 
\label{shotnoise5}
\end{eqnarray}
%%%%%%%%%%%%%%%%%%%%%%%%%%%%%%%%%%%%%%%%%%%%%%%%%%%%%%%%
holds at zero temperature. 
Our first aim in this paper is to show that $S=S_h$ holds 
up to ${\cal O}(V^3)$ by explicit calculations of 
the two quantities. 
Second aim is to calculate the Fano factor 
\cite{Sela06,Golub06,Gogolin06,Zarchin08}, 
%
%%%%%%%%%%%%%%%%%%%%%%%%%%%%%%%%%%%%%%%%%%%%%%%%%%%%%%%
\begin{eqnarray}
F_b = \frac{S_h}{2eI_b}, 
\label{shotnoise5}
\end{eqnarray}
%%%%%%%%%%%%%%%%%%%%%%%%%%%%%%%%%%%%%%%%%%%%%%%%%%%%%%%%
%
where the backscattering current $I_b$ is 
defined as $I_b =I_u-I$ with 
the current $I$ and the current 
in the unitarity limit $I_u$. 
\subsection{Renormalized perturbation theory}

For explicit calculations we employ 
the renormalized perturbation theory (RPT). 
It was introduced by Hewson\cite{HewsonRPT93,HewsonRPT04}
in equilibrium, and 
then extended by Oguri to the quantum dot 
system under a finite bias voltage\cite{OguriRPT05}. 
In this section the RPT applied 
for the Anderson Hamiltonian in eq.(\ref{dot}) 
is introduced first 
in equilibrium, and then out of equilibrium. 

The perturbation theory in 
$U$ for the Anderson model is 
characterized by a set of parameters: 
the energy level ${\epsilon}_d$, 
the resonance width $\Gamma$, 
and a regularized Coulomb 
interaction $U/\pi \Gamma$. 
In the RPT these parameters are substituted by renormalized ones, 
%%%%%%%%%%%%%%%%%%%%%%%%%%%%%%%%%%%%%%%%%%%%%%%%%%%%%%%
%%\begin{eqnarray}
$
{\epsilon}_d,{\ }\Gamma,{\ }U/\pi \Gamma \rightarrow 
\bar{\epsilon}_d,{\ }\bar{\Gamma},{\ }
\bar{U}/\pi \bar{\Gamma}. 
\label{renormalized}
$
%%\end{eqnarray}
%%%%%%%%%%%%%%%%%%%%%%%%%%%%%%%%%%%%%%%%%%%%%%%%%%%%%%%%
Using ward identities, three parameters are 
proved to be related with 
the regularized spin susceptibility 
$\tilde{\chi}_s \equiv [2\pi\Gamma/(g\mu_{\rm B})^2]\chi_s$, 
charge susceptibility 
$\tilde{\chi}_c \equiv (\pi\Gamma/2) \chi_c$
and 
specific heat coefficient 
$\tilde{\gamma} \equiv (3\Gamma/2\pi k_{\rm B}^2) \gamma$, 
%%%%%%%%%%%%%%%%%%%%%%%%%%%%%%%%%%%%%%%%%%%%%%%%%%%%%%%
\begin{eqnarray}
\nonumber
\tilde{\chi}_s &=& 
\bar{\rho}_{\sigma}(0) 
(1+ \bar{U} \bar{\rho}(0) ) \pi\Gamma, \\
\nonumber
\tilde{\chi}_c &=& \bar{\rho}(0) 
(1- \bar{U} \bar{\rho}(0) ) \pi\Gamma, \\
\tilde{\gamma} {\ }&=& \bar{\rho}(0) \pi\Gamma.
\label{th}
\end{eqnarray}
%%%%%%%%%%%%%%%%%%%%%%%%%%%%%%%%%%%%%%%%%%%%%%%%%%%%%%%%
The density of states at the Fermi level is given by 
%%%%%%%%%%%%%%%%%%%%%%%%%%%%%%%%%%%%%%%%%%%%%%%%%%%%%%%
\begin{eqnarray}
\bar{\rho}(0)=\frac{1}{\pi \bar{\Gamma}} 
\frac{1}{(\bar{\epsilon}_d /\bar{\Gamma})^2+1}. 
\label{dos}
\end{eqnarray}
%%%%%%%%%%%%%%%%%%%%%%%%%%%%%%%%%%%%%%%%%%%%%%%%%%%%%%%
Eq.(\ref{th}) leads to the Fermi liquid relation: 
$\tilde{\chi}_s + \tilde{\chi}_c=2 \tilde{\gamma}$, 
thus two quantities $\tilde{\chi}_s$ and $\tilde{\chi}_c$ 
become independent variables. 
The Wilson ratio defined as 
$R \equiv \tilde{\chi}_s/\tilde{\gamma}$ 
is expressed as 
$R = 2\tilde{\chi}_s/(\tilde{\chi}_s+\tilde{\chi}_c)$. 
As another relation, we consider 
the Friedel sum rule 
%%%%%%%%%%%%%%%%%%%%%%%%%%%%%%%%%%%%%%%%%%%%%%%%%%%%%%%
\begin{eqnarray}
\bar{n}=\frac{1}{2} -
\frac{1}{\pi} \tan^{-1} 
\left( \frac{\bar{\epsilon}_d}{\bar{\Gamma}} \right). 
\label{friedel}
\end{eqnarray}
%%%%%%%%%%%%%%%%%%%%%%%%%%%%%%%%%%%%%%%%%%%%%%%%%%%%%%%
Thus $\tilde{\chi}_s$, $\tilde{\chi}_c$ and $\bar{n}$ 
determine the three renormalized parameters
%%%%%%%%%%%%%%%%%%%%%%%%%%%%%%%%%%%%%%%%%%%%%%%%%%%%%%%
\begin{eqnarray}
\nonumber
\bar{\epsilon}_d&=&
%-2\sin n\pi \cos n\pi \Gamma 
-\sin (2\bar{n}\pi) \Gamma
/(\tilde{\chi}_s+\tilde{\chi}_c), \\
\nonumber
\bar{\Gamma}&=&
{\,}2{\,}\sin^2 (\bar{n}\pi) \Gamma 
/(\tilde{\chi}_s+\tilde{\chi}_c), \\
\bar{U}/\pi \bar{\Gamma}&=&
(R-1)/\sin^2 (\bar{n}\pi).
\label{rps}
\end{eqnarray}
%%%%%%%%%%%%%%%%%%%%%%%%%%%%%%%%%%%%%%%%%%%%%%%%%%%%%%%
From the bare Hamiltonian 
the renormalized parameters can be calculated 
using the exact Bethe ansatz results of 
$\tilde{\chi}_s$, $\tilde{\chi}_c$ and $\bar{n}$
\cite{BA}. 
Alternatively these parameters 
can be also estimated 
by the numerical renormalization group\cite{HewsonRPT04}. 

Here we comment on the behaviors of these parameters 
in the s-d limit and in the symmetric case. 

In the Kondo regime, 
according to Nozi$\grave{e}$res\cite{Nozieres74LFL}, 
we define the Kondo temperature 
as $T_{\rm K}= (g\mu_{\rm B})^2/\pi \chi_s$. 
The Bethe ansatz result of $\chi_s$ 
allows us to determine the expression of 
the Kondo temperature\cite{Hewsontxt}
%%%%%%%%%%%%%%%%%%%%%%%%%%%%%%%%%%%%%%%%%%%%%%%%%%%%%%%
\begin{eqnarray}
T_{\rm K} = \frac{4}{\pi} \sqrt{U\Gamma} 
\exp \left\{-\frac{\pi |\epsilon_d||\epsilon_d +U|}{2U\Gamma} 
 \right\}. 
\end{eqnarray}

%%%%%%%%%%%%%%%%%%%%%%%%%%%%%%%%%%%%%%%%%%%%%%%%%%%%%%%
In the Kondo limit, $\bar{n} \rightarrow 1/2$ 
and $\tilde{\chi}_c \rightarrow 0$, therefore 
$R=2$, $\bar{\epsilon}_d =0$, and 
$\pi \bar{\Gamma} = \bar{U}$. 
Moreover 
$\bar{\Gamma} = 2\Gamma /\tilde{\chi}_s = 
(g\mu_{\rm B})^2/\pi\chi_s$ is derived. 
Using the definition of the Kondo temperature, 
$\bar{\Gamma} =T_{\rm K}$ is obtained. 

The symmetric case gives rise to $\bar{n}=1/2$, 
thus $\bar{\epsilon}_d =0$, and 
%%%%%%%%%%%%%%%%%%%%%%%%%%%%%%%%%%%%%%%%%%%%%%%%%%%%%%%
\begin{eqnarray}
\frac{\bar{U}}{\pi \bar{\Gamma}} = R-1. 
\label{ugr-1}
\end{eqnarray}
%%%%%%%%%%%%%%%%%%%%%%%%%%%%%%%%%%%%%%%%%%%%%%%%%%%%%%%
Consequently physical quantities are 
expanded in a power series of $R-1$ 
for the symmetric case. 

As a check on RPT, 
the second-order correction in $\bar{U}$ 
to the self-energy has been calculated 
up to ${\cal O}(\omega^2)$ and ${\cal O}(T^2)$ 
for the symmetric case
\cite{HewsonRPT93}as 
%%%%%%%%%%%%%%%%%%%%%%%%%%%%%%%%%%%%%%%%%%%%%%%%%%%%%%%
\begin{eqnarray}
\bar{\Sigma}_d^r (\omega) 
= - i \frac{1}{2\bar{\Gamma}} 
\left( \frac{\bar{U}}{\pi \bar{\Gamma}} \right)^2 
(\omega^2 + \pi^2 T^2) + \cdots. 
\label{slf1}
\end{eqnarray}
%%%%%%%%%%%%%%%%%%%%%%%%%%%%%%%%%%%%%%%%%%%%%%%%%%%%%%%
This result is in agreement with the one 
in Ref.\onlinecite{YandY} 
where all orders in the bare $U$ are calculated 
at low frequencies and temperatures. 
Within ${\cal O}(\omega^2)$ and ${\cal O}(T^2)$, 
the second-order contribution in the renormalized 
$\bar{U}$ thus gives the exact expression. 
RPT enables us to discuss the exact Fermi-liquid features 
in the low energy region for all parameter regimes 
of ${\epsilon}_d$, $\Gamma$, and $U/\pi \Gamma$ 
from the weak correlation regime to 
the strong correlation regime. 

Having discussed RPT in equilibrium, 
let us turn to extension into 
under a finite bias voltage\cite{OguriRPT05}. 
Using the renormalized parameters, 
this procedure is done as follows. 

First we introduce Green functions 
based on Keldysh formalism
%%%%%%%%%%%%%%%%%%%%%%%%%%%%%%%%%%%%%%%%%%%%%%%%%%%%%%%
\begin{eqnarray}
\nonumber
& &\bar{G}_d(\tau-0)\equiv -i 
\langle T_c d_{\sigma}(\tau) d_{\sigma}^{\dagger}(0) \rangle, \\
\nonumber
& &\bar{G}_{pkd}(\tau-0)\equiv -i 
\langle T_c c_{pk\sigma}(\tau) d_{\sigma}^{\dagger}(0) \rangle, \\
\nonumber
& &\bar{G}_{dpk}(\tau-0)\equiv -i 
\langle T_c d_{\sigma}(\tau) c_{pk\sigma}^{\dagger}(0)  \rangle, \\
& &\bar{G}_{pkp'k'}(\tau-0)\equiv -i 
\langle T_c c_{pk\sigma}(\tau) c_{p'k'\sigma}^{\dagger}(0) \rangle,  
\label{g00}
\end{eqnarray}
%%%%%%%%%%%%%%%%%%%%%%%%%%%%%%%%%%%%%%%%%%%%%%%%%%%%%%%
where $\tau=t^{\pm}$ is the time variable. 
The lower and upper branch along the Keldysh contour 
are denoted by $-$ and $+$, and thus Keldysh components 
are defined as  
$A(t^{\alpha}-0^{\beta})\equiv A^{\alpha\beta}(t-0)$ 
for $\alpha,\beta=-,+$. 

The equation of motion enables us to 
relate these full Green functions with $\bar{G}_d$ 
%%%%%%%%%%%%%%%%%%%%%%%%%%%%%%%%%%%%%%%%%%%%%%%%%%%%%%%
\begin{eqnarray}
\nonumber
& &\bar{G}_d(\omega) = g_d^0(\omega) 
+ g_d^0(\omega) \bar{\Sigma}(\omega) \bar{G}_d(\omega) \\
\nonumber
& &\bar{G}_{pkd}(\omega) = \bar{V}_{pk} g_{pkpk}^0(\omega)
\sigma_z \bar{G}_d(\omega) \\
\nonumber
& &\bar{G}_{dpk}(\omega) = \bar{G}_{d}(\omega)
\sigma_z \bar{V}^*_{pk} g_{pkpk}^0(\omega) \\
\nonumber
& &\bar{G}_{pkp'k'}(\omega) = g_{pkp'k'}^0(\omega)
\delta_{pp'}\delta_{kk'} \\ 
& &{\quad}+ \bar{V}_{pk}g_{pkpk}^0(\omega) \sigma_z 
\bar{G}_d(\omega) \sigma_z \bar{V}^*_{p'k'}g_{p'k'p'k'}^0(\omega). 
%
%
%\end{array}
\label{g1}
\end{eqnarray}
%%%%%%%%%%%%%%%%%%%%%%%%%%%%%%%%%%%%%%%%%%%%%%%%%%%%%%%
$\bar{G}$ represents the matrix 
$(\bar{G})^{\alpha\beta}\equiv \bar{G}^{\alpha\beta}$ and 
$\sigma_z$ is the third Pauli matrix in the 
$\alpha\beta$ space.
Here $g^0$ refers to $U=0$ and $V_{L,Rk}=0$. 

As the zeroth-order Green functions in $\bar{U}$,
we consider the initial Green functions 
$\bar{g}  \equiv \bar{G}|_{\bar{U}=0} $, 
renormalized by $\bar{V}_{L,Rk}$. 
The initial Green functions are obtained as 
the solution of the Dyson eq. for 
$\bar{G}_d|_{\bar{U}=0} \equiv \bar{g}_d$ in eq.(\ref{g1}), 
%%%%%%%%%%%%%%%%%%%%%%%%%%%%%%%%%%%%%%%%%%%%%%%%%%%%%%%
\begin{eqnarray}
\nonumber
& &\bar{g}^{--}_d(\omega) =
(1-f_{\rm eff}) \bar{g}^r_d(\omega) + 
 f_{\rm eff} \bar{g}^a_d(\omega)  \\
\nonumber
& &\bar{g}^{-+}_d(\omega)=
-f_{\rm eff} (\bar{g}^r_d(\omega) 
- \bar{g}^a_d(\omega))  \\
\nonumber
& &\bar{g}^{+-}_d(\omega) =
(1-f_{\rm eff}) (\bar{g}^r_d(\omega) 
- \bar{g}^a_d(\omega))  \\
& &\bar{g}^{++}_d(\omega) =
-(1-f_{\rm eff}) \bar{g}^a_d(\omega) - 
 f_{\rm eff} \bar{g}^r_d(\omega).  
%\end{array}
\label{g0}
\end{eqnarray}
%%%%%%%%%%%%%%%%%%%%%%%%%%%%%%%%%%%%%%%%%%%%%%%%%%%%%%%
$\bar{g}^r_{\sigma}(\omega)=1/(\omega -\bar{\epsilon}_d 
+i \bar{\Gamma})$, 
$\bar{g}^a_{\sigma}(\omega)=g^r(\omega)_{\sigma}^*$, 
%%%%%%%%%%%%%%%%%%%%%%%%%%%%%%%%%%%%%%%%%%%%%%%%%%%%%%%
\begin{eqnarray}
f_{\rm eff} =
\frac{\bar{\Gamma}_L f_L + \bar{\Gamma}_R f_R }
{\bar{\Gamma}_L+\bar{\Gamma}_R}, {\ \ }
\end{eqnarray}
%%%%%%%%%%%%%%%%%%%%%%%%%%%%%%%%%%%%%%%%%%%%%%%%%%%%%%%
Here the Fermi distribution function 
$f_{L,R} = 1/(1+{\rm e}^{\beta(\omega \mp eV/2)})$
are not renormalized. 
Substituting 
$\bar{G}_d|_{U=0}=\bar{g}_d$ into eq.(\ref{g1}), 
we get the explicit forms of 
other initial Green functions
$\bar{g}_{pkd}$, $\bar{g}_{dpk}$ and $\bar{g}_{pkp'k'}$. 
In the perturbation theory in $\bar{U}$, 
we use these initial Green functions $\bar{g}$. 
%%%%%%%%%%%%%%%%%%%%%%%%%%%%%%%%%%%%%%%%%%%%%%%%%%%%%%%
%%\begin{eqnarray}
%%\nonumber
%%& &\Sigma(\omega) \equiv \Sigma_L(\omega)+\Sigma_R(\omega) \\
%%& &\Sigma_{L,R}(\omega)_p= 
%%\sigma_z \left( 
%%\sum_k|V_{L,Rk}|^2 g_{pkpk}^0(\omega)
%%\right) \sigma_z
%%\label{s1}
%%\end{eqnarray}
%%%%%%%%%%%%%%%%%%%%%%%%%%%%%%%%%%%%%%%%%%%%%%%%%%%%%%%

Using the initial Green function $\bar{g}_d$, 
the Dyson equation of $\bar{G}_d$ in eq.(\ref{g1}) 
is rewritten into 
%%%%%%%%%%%%%%%%%%%%%%%%%%%%%%%%%%%%%%%%%%%%%%%%%%%%%%%
\begin{eqnarray}
\bar{G}_d(\omega) = \bar{g}_d(\omega) 
+ \bar{g}_d(\omega) \bar{\Sigma}_d(\omega) \bar{G}_d(\omega). 
%%
%%\bar{G}^{\alpha \beta}_{\sigma}(\omega)=
%%\bar{g}^{\alpha \beta}_{\sigma}(\omega)
%%+\bar{g}^{\alpha \gamma}_{\sigma}(\omega)
%%\bar{\Sigma}^{\gamma \delta}_{\sigma}(\omega)
%%\bar{G}^{\delta \beta}_{\sigma}(\omega),
\label{dyson}
\end{eqnarray}
%%%%%%%%%%%%%%%%%%%%%%%%%%%%%%%%%%%%%%%%%%%%%%%%%%%%%%%
With a unitary transformation, 
eq.(\ref{dyson}) is reduced to the three-components form. 
One of them becomes the Dyson equation for the retarded 
component, 
%%%%%%%%%%%%%%%%%%%%%%%%%%%%%%%%%%%%%%%%%%%%%%%%%%%%%%%
\begin{eqnarray}
\bar{G}^r_d(\omega)=\bar{g}^r_d(\omega) + 
\bar{g}^r_d(\omega) \bar{\Sigma}_d^r (\omega)
\bar{G}^r_d(\omega). 
\label{rdyson}
\end{eqnarray}
%%%%%%%%%%%%%%%%%%%%%%%%%%%%%%%%%%%%%%%%%%%%%%%%%%%%%%%
The current is expressed as the renormalized form, 
%%%%%%%%%%%%%%%%%%%%%%%%%%%%%%%%%%%%%%%%%%%%%%%%%%%%%%%
\begin{eqnarray}
I=\frac{e}{h} \sum_{\sigma} \int^{\infty}_{-\infty} d\omega
\bar{T}(\omega) 
(f_L-f_R), 
\label{current}
\end{eqnarray}
%%%%%%%%%%%%%%%%%%%%%%%%%%%%%%%%%%%%%%%%%%%%%%%%%%%%%%%
where transmission probability $\bar{T}(\omega)$ is
defined as 
%%%%%%%%%%%%%%%%%%%%%%%%%%%%%%%%%%%%%%%%%%%%%%%%%%%%%%%

\begin{eqnarray}
\bar{T}(\omega) = 
\frac{\bar{\Gamma}_L \bar{\Gamma}_R}
{\bar{\Gamma}_L +\bar{\Gamma}_R}
\bar{\rho}(\omega), {\ \ \ }
\rho(\omega)= -\frac{1}{\pi} 
{\rm Im} \bar{G}^r_d(\omega)
\label{trans}
\end{eqnarray}
%%%%%%%%%%%%%%%%%%%%%%%%%%%%%%%%%%%%%%%%%%%%%%%%%%%%%%%

Under a finite bias voltage, the second-order calculation 
in $\bar{U}$ for $\bar{\Sigma}^r$ has 
been done for the symmetric case\cite{OguriRPT05}, 
%%%%%%%%%%%%%%%%%%%%%%%%%%%%%%%%%%%%%%%%%%%%%%%%%%%%%%%
\begin{eqnarray}
\bar{\Sigma}^r_d (\omega) 
= - i \frac{1}{2\bar{\Gamma}} 
\left( \frac{\bar{U}}{\pi \bar{\Gamma}} \right)^2 
(\omega^2 + (eV)^2 + \pi^2 T^2) + \cdots. 
\label{slf2}
\end{eqnarray}
%%%%%%%%%%%%%%%%%%%%%%%%%%%%%%%%%%%%%%%%%%%%%%%%%%%%%%%
Clearly this corresponds to eq.(\ref{slf1}) 
in the equilibrium limit $V=0$ as expected, 
and moreover reproduces the complete result 
if combined with the Ward identity\cite{Oguri01Ward}. 

We substitute $\bar{\Sigma}^r (\omega)$ 
of eq.(\ref{slf2}) to 
the formal solution of $G^r$ in eq.(\ref{rdyson}), 
and calculate the transmission 
probability with eq.(\ref{trans})
%%%%%%%%%%%%%%%%%%%%%%%%%%%%%%%%%%%%%%%%%%%%%%%%%%%%%%%
\begin{eqnarray}
\bar{T}(\omega)
= 1- \frac{\omega^2}{\bar{\Gamma}^2}
-\frac{1}{2\bar{\Gamma}^2} 
\left( \frac{\bar{U}}{\pi\bar{\Gamma}} \right)^2
\left\{ \omega^2  
 + 
\frac{3}{4} (eV)^2 
\right\} +\cdots
\label{tex}
\end{eqnarray}
%%%%%%%%%%%%%%%%%%%%%%%%%%%%%%%%%%%%%%%%%%%%%%%%%%%%%%%
The resulting transmission probability 
allows us to calculate the current defined 
in eq.(\ref{current}). 
For later use, we show the current 
at zero temperature 
%%%%%%%%%%%%%%%%%%%%%%%%%%%%%%%%%%%%%%%%%%%%%%%%%%%%%%%
\begin{eqnarray}
I&=&\frac{2e^2}{h} V
\left\{ 1-\frac{1+5(R-1)^2}{12} 
\left( \frac{eV}{\bar{\Gamma}} \right)^2 \right\} 
+ \cdots,
\label{current2}
\\
&\rightarrow& \frac{2e^2}{h} V
\left\{ 1-\frac{1}{2} 
\left( \frac{eV}{T_{\rm K}} \right)^2 \right\} 
+ \cdots,
{\ \ {\rm s-d{\ }limit}}
\label{sdcurrent2}
\end{eqnarray}
%%%%%%%%%%%%%%%%%%%%%%%%%%%%%%%%%%%%%%%%%%%%%%%%%%%%%%%
where $\bar{U}/\pi\bar{\Gamma}=R-1$ is used in 
eq.(\ref{current2}). 
In the Kondo limit, $R=2$ and $\bar{\Gamma}=T_{\rm K}$. 
The s-d limit result given in eq.(\ref{sdcurrent2}) 
completely agrees with the ones 
in Refs.\onlinecite{Glatxmantxt04,Sela06,Golub06,Gogolin06} 
where the fixed point Hamiltonian are used. 

Therefore, for the quantum dot system 
the second-order calculations in $\bar{U}$
to the self-energy can provide the exact result 
up to 
${\cal O}(\omega^2)$, ${\cal O}(T^2)$ and ${\cal O}(V^2)$. 
This result leads to 
the correct expression of current up to 
${\cal O}(V^3)$, where we need not count 
higher-order corrections in $\bar{U}$. 

\subsection{Calculation of $S$ and $S_h$}

In this section, we consider 
shot noise for the symmetric case 
at zero temperature up to ${\cal O}(V^3)$,
by employing the two definitions of the 
shot noise, 
noise power $S$ at $T=0$, and 
the new formula for shot noise $S_h$. 
Within ${\cal O}(V^3)$, in the same way 
as the current, 
the second-order calculations in $\bar{U}$ 
are sufficient to give 
correct results for $S$ and $S_h$. 

Both the noise power $S$ in eq.(\ref{npdf}) 
and the shot noise $S_h$ in eq.(\ref{shdf}) 
are two-particle Green functions. 
Generally they are expressed by 
the bubble-type diagrams (bubble-diagrams) 
and the vertex-correction-type diagrams 
(vertex-diagrams). 
We use subscripts $1$ and $2$ to 
denote contributions from 
the bubble-diagrams and vertex-diagrams. 
Thus $S=S_1+S_2$ and $S_h=S_{h1}+S_{h2}$. 
%
%%%%%%%%%%%%%%%%%%%%%%%%%%%%%%%%%%%%%%%%%%%%%%%%%%%%%%%%%%%%
\begin{figure}[h]
\includegraphics[width=7cm]{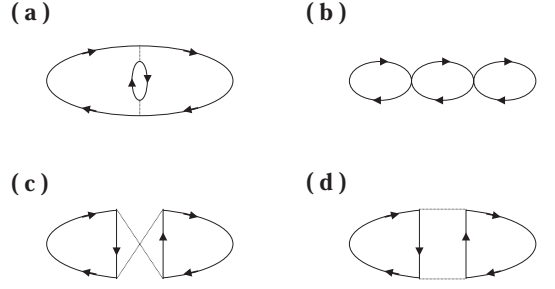}
\caption{\label{1} For the symmetric case, 
the second-order corrections to 
the vertex-diagrams.}
\label{diagram}
\end{figure}
%
%%%%%%%%%%%%%%%%%%%%%%%%%%%%%%%%%%%%%%%%%%%%%%%%%%%%%%%%%%%%%
%
%
\subsubsection{Noise power $S$}
This section is devoted to the analysis of noise power $S$. 
As mentioned previously, 
RPT in $\bar{U}$ based on Keldysh formalism is employed. 
We begin with giving a definition of $S$ 
in the Keldysh form. 
Considering the steady-state feature in eq.(\ref{st}), 
the original $S$ in eq.(\ref{npdf}) 
expressed by the anticommutation relation 
is rewritten into 
%%%%%%%%%%%%%%%%%%%%%%%%%%%%%%%%%%%%%%%%%%%%%%%%%%%%%%%
\begin{eqnarray}
\nonumber
S&=& 2 \int^{\infty}_{-\infty}dt 
\langle {\;}T_c{\;}J_H(t^+)J_H(0^-) \rangle_{L} \\
&=& \frac{e^2}{\hbar} 
\left\{ 2{\;}{\rm Re}(K^{+-}_a-K^{+-}_b) \right\}. 
\label{np0}
\end{eqnarray}
%%%%%%%%%%%%%%%%%%%%%%%%%%%%%%%%%%%%%%%%%%%%%%%%%%%%%%%%
$\langle ... \rangle_{L}$ means the linked parts. 
$T_c$ is the time-ordering operator defined on 
the Keldysh contour. The time variable $t^{\pm}$ 
represent times on the upper and lower branches. 
It is convenient to classify 
the current-current correlation functions 
into $K^{+-}_{a,b}$ 
%%%%%%%%%%%%%%%%%%%%%%%%%%%%%%%%%%%%%%%%%%%%%%%%%%%%%%%
\begin{eqnarray}
K^{+-}_{a,b} = \int dt {\;}K^{+-}_{a,b}(t-0), 
\label{np1}
\end{eqnarray}
%%%%%%%%%%%%%%%%%%%%%%%%%%%%%%%%%%%%%%%%%%%%%%%%%%%%%%%%
where $K^{+-}_{a,b}(t-0) \equiv K_{a,b}(t^+ -0^-)$. 
Here 
%%%%%%%%%%%%%%%%%%%%%%%%%%%%%%%%%%%%%%%%%%%%%%%%%%%%%%%
\begin{eqnarray}
\nonumber
  K_a (\tau-0){\!\!}&=&  (i^2/2) 
\sum_{pp'} \sigma_z^{pp} \sigma_z^{p'p'} 
\sum_{kk'\sigma\sigma'} 
\bar{V}_{pk} \bar{V}_{p'k'} \\
\nonumber
&\times& {\!\!\!}
\langle {\;}T_c{\;}
  c^{\dagger}_{Hpk\sigma}(\tau) d_{H\sigma}(\tau) 
 \ c^{\dagger}_{Hp'k'\sigma'}(0) d_{H\sigma'}(0)
 \rangle_L ,  \\
\nonumber
  K_b (\tau-0){\!\!}&=&  (i^2/2) 
\sum_{pp'} \sigma_z^{pp} \sigma_z^{p'p'}
\sum_{kk'\sigma\sigma'} 
\bar{V}_{pk} \bar{V}^{*}_{p'k'} \\
\nonumber
&\times& {\!\!\!}
\langle {\;}T_c{\;}
  c^{\dagger}_{Hpk\sigma}(\tau) d_{H\sigma}(\tau) 
 \ d^{\dagger}_{H\sigma'}(0) c_{Hp'k'\sigma'}(0) 
 \rangle_L  
\label{np2}
\end{eqnarray}
%%%%%%%%%%%%%%%%%%%%%%%%%%%%%%%%%%%%%%%%%%%%%%%%%%%%%%%%

$K_{a,b}(\tau-0)$ are also the two-particle 
Green functions. Thus $K_{a,b}$ is generally 
given by contributions of 
the bubble-diagrams $K_{1a,b}$ and 
the vertex-diagrams $K_{2a,b}$ as 
$K_{a,b}=K_{1a,b}+K_{2a,b}$. 
Appling this relation for $S$ 
in eq.(\ref{np0}), we define 
$S_{1,2}$ to satisfy $S=S_1+S_2$ as follows, 
%%%%%%%%%%%%%%%%%%%%%%%%%%%%%%%%%%%%%%%%%%%%%%%%%%%%%%%
\begin{eqnarray}
\nonumber
S_1&=& \frac{e^2}{\hbar} 
\left\{ 2{\;}{\rm Re}(K^{+-}_{1a}-K^{+-}_{1b}) \right\} \\
S_2&=& \frac{e^2}{\hbar} 
\left\{ 2{\;}{\rm Re}(K^{+-}_{2a}-K^{+-}_{2b}) \right\}. 
\label{np012}
\end{eqnarray}
%%%%%%%%%%%%%%%%%%%%%%%%%%%%%%%%%%%%%%%%%%%%%%%%%%%%%%%%
Thus, explicit calculations of $K_{1,2a,b}$ 
determine noise power $S$. 

Let us start with the discussion of $S_1$. 
$K_{1,a,b}$ are evaluated as 
%%%%%%%%%%%%%%%%%%%%%%%%%%%%%%%%%%%%%%%%%%%%%%%%%%%%%%%
\begin{eqnarray}
\nonumber
  K_{1a} (\tau-0)&=&  (i^2/2) 
\sum_{pp'} \sigma_z^{pp} \sigma_z^{p'p'}
\sum_{kk'\sigma\sigma'} 
\bar{V}_{pk} \bar{V}_{p'k'} \\
\nonumber
&\times& 
\bar{G}_{dp'k'}(\tau-0) \bar{G}_{dpk}(0-\tau)
{\;}\delta_{\sigma\sigma'},  \\
\nonumber
  K_{1b} (\tau-0)&=&  (i^2/2) 
\sum_{pp'} \sigma_z^{pp} \sigma_z^{p'p'}
\sum_{kk'\sigma\sigma'} 
\bar{V}_{pk} \bar{V}^{*}_{p'k'} \\
&\times& 
\bar{G}_{d}(\tau-0) \bar{G}_{p'k'pk}(0-\tau).
{\;}\delta_{\sigma\sigma'}
\label{np3}
\end{eqnarray}
%%%%%%%%%%%%%%%%%%%%%%%%%%%%%%%%%%%%%%%%%%%%%%%%%%%%%%%%
In the following, we use the Fourier representation. 

It is shown that $\bar{G}_{dpk}$ 
and $\bar{G}_{p'k'pk}$ 
are related with $\bar{G}_{d}$ in eq.(\ref{g1}). 
Technically, these relations are sufficient 
to proceed further calculations. 
However, here it is better to comment on 
a difference between calculations of 
noise power and shot noise. 

Here we concern with the $k$-sum in eq.(\ref{np3}). 
In both cases 
the $k$-sum is reduced to a form 
%%%%%%%%%%%%%%%%%%%%%%%%%%%%%%%%%%%%%%%%%%%%%%%%%%%%%%%
\begin{eqnarray}
\nonumber
{\ }
\sum_k \bar{V}_{pk} \bar{G}_{dpk}{\ },
\sum_{kk'} \bar{V}_{pk} \bar{V}^{*}_{p'k'}
\bar{G}_{p'k'pk} 
\rightarrow 
\sum_{k}|V_{pk}|^2 g^0_{pkpk}(\omega).
\label{npk}
\end{eqnarray}
%%%%%%%%%%%%%%%%%%%%%%%%%%%%%%%%%%%%%%%%%%%%%%%%%%%%%%%%
$g^0_{pkpk}(\omega)$ are the 
bare Green functions for the free electrons 
in leads when $U=0$ and $V_{pk}=0$. 
Thus the analytic properties of $g^0_{pkpk}(\omega)$ 
are characterized by infinitesimal quantities 
$\pm i\epsilon$. 
By performing the $k$-sum of $g^0_{pkpk}$, 
the $\delta$-function singularities are averaged. 
Therefore in the full bubble-diagram calculations, 
the $k$-sum provides well-defined quantities. 

However, in the shot-noise calculations, 
another type of the $k$-sum 
leads to a $\delta$-function singularity. 
We will discuss this point in the next section. 
%%%%%%%%%%%%%%%%%%%%%%%%%%%%%%%%%%%%%%%%%%%%%%%%%%%%%%%
%%\begin{eqnarray}
%%%
%%\nonumber
%%& &\sum_{k} 
%%\bar{V}_{pk} \bar{G}_{pkd}(\omega) = 
%%\left(
%%\sum_{k} |\bar{V}_{pk}|^2 g_{pkpk}^0(\omega) \right)
%%\sigma_z \bar{G}_d(\omega) \\
%%%
%%\nonumber
%%& &\bar{G}_{dpk}(\omega) = \bar{G}_{d}(\omega)
%%\sigma_z \bar{V}^*_{pk} g_{pkpk}^0(\omega) \\
%%%
%%\nonumber
%%& &
%%\sum_{kk'} 
%%\bar{V}_{pk}\bar{V}^*_{p'k'} \bar{G}_{pkp'k'}(\omega) 
%%= \sum_{k} |\bar{V}_{pk}|^2 g_{pkp'k}^0(\omega)

%%\delta_{pp'} \\ 
%%& &{\quad}+ 
%%\left( \sum_{k} |\bar{V}_{pk}|^2 g_{pkpk}^0(\omega) \right)
%%\sigma_z \bar{G}_d(\omega) \sigma_z 
%%\left(
%%\sum_{k'} |\bar{V}_{p'k'}|^2 g_{p'k'p'k'}^0(\omega) \right). 
%\end{array}
%%\label{g1}
%%\end{eqnarray}
%%%%%%%%%%%%%%%%%%%%%%%%%%%%%%%%%%%%%%%%%%%%%%%%%%%%%%%

Let us return back to the main discussion. 
The relations in eq.(\ref{g1}) allows us to 
rewrite $2{\;}{\rm Re}(K^{+-}_{1a}-K^{+-}_{1b})$ 
into an expression with $\bar{T}$, 
%%%%%%%%%%%%%%%%%%%%%%%%%%%%%%%%%%%%%%%%%%%%%%%%%%%%%%%
\begin{eqnarray}
\nonumber
2{\;}{\rm Re}(K_{1a}-K_{1b})=
\int \frac{d\omega}{2\pi} 
\left\{ -4 \bar{T}^2 (\omega) (f_L-f_R)^2 \right. \\
+4 \bar{T} (\omega) (f_L (1-f_R)+f_R(1-f_L)). 
\label{np4}
\end{eqnarray}
%%%%%%%%%%%%%%%%%%%%%%%%%%%%%%%%%%%%%%%%%%%%%%%%%%%%%%%%
Using 
%%%%%%%%%%%%%%%%%%%%%%%%%%%%%%%%%%%%%%%%%%%%%%%%%%%%%%%%
\begin{eqnarray}
\nonumber
& &f_L (1-f_R)+f_R(1-f_L)= \\
& &{\quad}f_L (1-f_L)+f_R(1-f_R)+(f_L-f_R)^2, 
\label{np5}
\end{eqnarray}
%%%%%%%%%%%%%%%%%%%%%%%%%%%%%%%%%%%%%%%%%%%%%%%%%%%%%%%%
$S_1$ can be represented as 
%%%%%%%%%%%%%%%%%%%%%%%%%%%%%%%%%%%%%%%%%%%%%%%%%%%%%%%
\begin{eqnarray}
\nonumber
S_1=
\frac{4e^2}{h} 
\int d\omega \bar{T} (\omega) 
(f_L (1-f_L)+f_R(1-f_R)) & &\\
+\frac{4e^2}{h} 
\int d\omega \bar{T} (\omega) (1-\bar{T} (\omega))
(f_L-f_R)^2.& & 
\label{s1}
\end{eqnarray}
%%%%%%%%%%%%%%%%%%%%%%%%%%%%%%%%%%%%%%%%%%%%%%%%%%%%%%%%
We have proven that noise power $S_1$ for 
the bubble-diagrams 
decouples into the thermal-noise part and shot-noise part
in the same manner as noninteracting systems
\cite{beenakker,blanter}. 
However, in this case the transmission probability 
is fully renormalized by the Coulomb interaction. 

At zero temperature, in $S_1$ the thermal-noise part 
vanishes. 
Consequently we discuss 
the remaining shot-noise part 
in the second-line of eq.(\ref{s1}). 
With $\bar{T}$ given by eq.(\ref{tex}), 
we evaluate the asymptotic form of $S_1$ 
%%%%%%%%%%%%%%%%%%%%%%%%%%%%%%%%%%%%%%%%%%%%%%%%%%%%%%%
\begin{eqnarray}
S_{1} =
\frac{4e^3}{h} |V| 
\left\{
\frac{1}{12} \left( \frac{eV}{\bar{\Gamma}} \right)^2
{\!\!\!\!}
+ \frac{5}{12}(R-1)^2 \left( \frac{eV}{\bar{\Gamma}} \right)^2
\right\},
\label{as1}
\end{eqnarray}
%%%%%%%%%%%%%%%%%%%%%%%%%%%%%%%%%%%%%%%%%%%%%%%%%%%%%%%%
where $\bar{U}/\pi\bar{\Gamma}=R-1$. 

Here, let us turn to the discussion of $S_2$ 
for the vertex-diagrams. Diagrams 
up to the second order in $\bar{U}$ 
are shown in Fig.~\ref{diagram}. 
For the symmetric case, 
we can show that the contribution from the diagram 
in (c) cancels out the one from (d), and the 
one from (b) itself 
vanishes. 
We explicitly calculate the remaining diagram in (a). 

We calculate $K_{2a}$ 
up to the second order in $\bar{U}$ 
%%%%%%%%%%%%%%%%%%%%%%%%%%%%%%%%%%%%%%%%%%%%%%%%%%%%%%%
\begin{eqnarray}
K_{2a,b}(\tau -0)&=& 2 \int d\tau_1 d\tau_2 F_{a,b}, 
\label{np6}
\end{eqnarray}
%%%%%%%%%%%%%%%%%%%%%%%%%%%%%%%%%%%%%%%%%%%%%%%%%%%%%%%
where 
%%%%%%%%%%%%%%%%%%%%%%%%%%%%%%%%%%%%%%%%%%%%%%%%%%%%%%%
\begin{eqnarray}
\nonumber
F_a &=&
\bar{g}_d(\tau-\tau_1) 
\sum_{p'}
 \sigma_z^{p'p'} \sum_{k'} \bar{V}_{p'k'} 
\bar{g}_{dp'k'}(\tau_1-0) \\
\nonumber
&\times& 
\bar{g}_d(0-\tau_2)
\sum_{p} \sigma_z^{pp} 
\sum_{k} \bar{V}_{pk} \bar{g}_{dpk}(\tau_2-\tau) \\
& & {\qquad\qquad\qquad}
\times \bar{U}\bar{g}_d(\tau_1-\tau_2) 
\bar{U}\bar{g}_d(\tau_2-\tau_1), \\
\nonumber
F_b &=&
\bar{g}_d(\tau-\tau_1) 
\bar{g}_{d}(\tau_1-0) \times \sum_{pp'} \\
\nonumber
& &
\sum_{k'} \bar{V}^*_{p'k'} \bar{g}_{p'k'd}(0-\tau_2) 
\sigma_z^{p'p} 
\sum_{k} \bar{V}_{pk} \bar{g}_{pkd}(\tau_2-\tau) \\
& &{\qquad\qquad\qquad}
\times \bar{U}\bar{g}_d(\tau_1-\tau_2) \bar{U}
\bar{g}_d(\tau_2-\tau_1). 
\label{np7}
\end{eqnarray}
%%%%%%%%%%%%%%%%%%%%%%%%%%%%%%%%%%%%%%%%%%%%%%%%%%%%%%%
With eq.(\ref{np012})
$S_2$ is obtained in the Fourier 
representation
%%%%%%%%%%%%%%%%%%%%%%%%%%%%%%%%%%%%%%%%%%%%%%%%%%%%%%%
\begin{eqnarray}
\nonumber
S_{2}=
\frac{4e^2}{\hbar} c 
\int d\omega_{1}d\omega_{2} {\;}F_{ab},
\label{np8}
\end{eqnarray}
%%%%%%%%%%%%%%%%%%%%%%%%%%%%%%%%%%%%%%%%%%%%%%%%%%%%%%%
where $c=\bar{U}^2\bar{\Gamma}^2/(2\pi)^2$. 
$F_{ab}$ is given by 
%%%%%%%%%%%%%%%%%%%%%%%%%%%%%%%%%%%%%%%%%%%%%%%%%%%%%%%
\begin{eqnarray}
\nonumber
F_{ab}=
\sum_{\alpha}
g_d(\omega_1)^{\alpha -} 
\sigma_z^{\alpha\bar{\alpha}}g_d(\omega_1)^{+\bar{\alpha}}
( f_L-f_R )({\omega_1 +\omega_2}) & \\
\nonumber
{\ }\times 
g_d(\omega_2)^{-\alpha }g_d(\omega_2)^{\bar{\alpha}+}
( f_L-f_R )({\omega_2}) 
\times
\Gamma(\omega_2-\omega_1), &
\label{np9}
\end{eqnarray}
%%%%%%%%%%%%%%%%%%%%%%%%%%%%%%%%%%%%%%%%%%%%%%%%%%%%%%%
where $\bar{\alpha}\equiv-\alpha$ and 
$( f_L-f_R )({\omega})\equiv f_L(\omega)-f_R(\omega)$. 
The vertex function $\Gamma(\omega)$ is defined as 
%%%%%%%%%%%%%%%%%%%%%%%%%%%%%%%%%%%%%%%%%%%%%%%%%%%%%%%
\begin{eqnarray}
\nonumber
&&\Gamma(\omega) 
= 
\int d\omega' g_d^{-+}(\omega') g_d^{+-}(\omega+\omega') \\
\nonumber
&&{\ }
\simeq \theta (\omega) \frac{2\omega}{\bar{\Gamma}^2}
+\theta (\omega -eV) \frac{\omega -eV}{\bar{\Gamma}^2}
+\theta (\omega +eV) \frac{\omega +eV}{\bar{\Gamma}^2}. \\
\label{np10}
\end{eqnarray}
%%%%%%%%%%%%%%%%%%%%%%%%%%%%%%%%%%%%%%%%%%%%%%%%%%%%%%%
Performing the frequency-integral gives 
the explicit form of $\Gamma(\omega)$. 
In eq.(\ref{np10}) we show the leading-order 
of $\Gamma(\omega)$. 

Finally $S_2$  becomes  
%%%%%%%%%%%%%%%%%%%%%%%%%%%%%%%%%%%%%%%%%%%%%%%%%%%%%%%
\begin{eqnarray}
S_2= \frac{4e^2}{h}|V|
\left\{ \frac{1}{3}(R-1)^2 
\left( \frac{eV}{\bar{\Gamma}} \right)^2 \right\}.  
+ \cdots
\label{as2}
\end{eqnarray}
%%%%%%%%%%%%%%%%%%%%%%%%%%%%%%%%%%%%%%%%%%%%%%%%%%%%%%%%

Therefore $S=S_1+S_2$ is also obtained 
up to the leading-order in $V$, 
%%%%%%%%%%%%%%%%%%%%%%%%%%%%%%%%%%%%%%%%%%%%%%%%%%%%%%%
\begin{eqnarray}
\nonumber
& &
{\!\!\!\!\!\!\!\!}S =
\frac{4e^3}{h} |V| 
\left\{
\frac{1}{12} \left( \frac{eV}{\bar{\Gamma}} \right)^2 \right.
{\!\!\!\!\!}
+ \frac{5}{12}(R-1)^2 
\left( \frac{eV}{\bar{\Gamma}} \right)^2  \\
%
%\nonumber
& & {\qquad\qquad\qquad\qquad\qquad} 
{\!\!\!\!\!\!}
+\left. {\,}\frac{1}{3}{\,}(R-1)^2 
\left( \frac{eV}{\bar{\Gamma}} \right)^2
\right\}.  
\label{as}
\end{eqnarray}
%%%%%%%%%%%%%%%%%%%%%%%%%%%%%%%%%%%%%%%%%%%%%%%%%%%%%%%%
%
%
\subsubsection{Shot noise $S_h$}

This section is devoted to calculations of the new 
formula of shot noise 
%%%%%%%%%%%%%%%%%%%%%%%%%%%%%%%%%%%%%%%%%%%%%%%%%%%%%%%
\begin{eqnarray}
S_h= 
-\langle \{ \delta J,e(\delta N_L -\delta N_R) \} \rangle .
\label{sh1}
\end{eqnarray}
%%%%%%%%%%%%%%%%%%%%%%%%%%%%%%%%%%%%%%%%%%%%%%%%%%%%%%%%
Originally, $S_h$ contains 
$4\times4=16$ two-particle Green functions. 
The current conservation $\langle J_L+J_R \rangle =0$ 
reduces the number to 8 two-particle Green functions. 
As a preliminary step, we define the reduced $S_h$ 
in the Keldysh form
%%%%%%%%%%%%%%%%%%%%%%%%%%%%%%%%%%%%%%%%%%%%%%%%%%%%%%%
\begin{eqnarray}
S_h =
\frac{e^2}{\hbar} 2{\;} {\rm Re} \left\{
(K_{LL} + K_{RR})- (K_{LR} +K_{RL})\right\}, 
\label{sh2}
\end{eqnarray}
%%%%%%%%%%%%%%%%%%%%%%%%%%%%%%%%%%%%%%%%%%%%%%%%%%%%%%%
where $K_{pp'}$ for ${p,p'}={L,R}$ are given by 
%%%%%%%%%%%%%%%%%%%%%%%%%%%%%%%%%%%%%%%%%%%%%%%%%%%%%%%
\begin{eqnarray}
\nonumber
 & &K_{pp'} = -i \sum_{kk'\sigma\sigma'} V_{pk}  \\
\nonumber
& &\times 
\langle {\;}T_c{\;}
  c^{\dagger}_{Hpk\sigma}(0^+) d_{H\sigma}(0^+) 
 \ c^{\dagger}_{Hp'k'\sigma'}(0^-) c_{Hp'k'\sigma'}(0^-)
 \rangle_L .
\label{sh3}
\end{eqnarray}
%%%%%%%%%%%%%%%%%%%%%%%%%%%%%%%%%%%%%%%%%%%%%%%%%%%%%%%%
The equal-time correlation functions are 
defined by using the Keldysh branches. 

$K_{pp'}$ as the two-particle Green functions 
are generally given 
by $K_{1pp'}$ for the bubble-diagrams and 
$K_{2pp'}$ for the vertex-diagrams.  
Correspondingly, $S_{h1,2}$ are defined 
as $S_h = S_{h1}+S_{h2}$: 
%%%%%%%%%%%%%%%%%%%%%%%%%%%%%%%%%%%%%%%%%%%%%%%%%%%%%%%
\begin{eqnarray}
\nonumber
S_{h1} &=&
\frac{e^2}{\hbar} 2{\;} {\rm Re} \left\{
(K_{1LL} + K_{1RR})- (K_{1LR} +K_{1RL})\right\}, \\
\nonumber
S_{h2} &=&
\frac{e^2}{\hbar} 2{\;} {\rm Re} \left\{
(K_{2LL} + K_{2RR})- (K_{2LR} +K_{2RL})\right\}. \\
\label{sh20}
\end{eqnarray}
%%%%%%%%%%%%%%%%%%%%%%%%%%%%%%%%%%%%%%%%%%%%%%%%%%%%%%%
Therefore an evaluation of $K_{1,2pp'}$ 
determines $S_h$. 

First we discuss $K_{1pp'}$
%%%%%%%%%%%%%%%%%%%%%%%%%%%%%%%%%%%%%%%%%%%%%%%%%%%%%%%
\begin{eqnarray}
\nonumber
{\;\;}
K_{1pp'} {\!\!\!}&=& {\!\!\!}
-i \sum_{kk'\sigma\sigma'} 
\delta_{\sigma\sigma'}
\bar{V}_{pk}  
\bar{G}_{p'k'pk}^{-+}(0) \bar{G}_{dp'k'}^{+-}(0), \\
\nonumber
&=& {\!\!\!}
-i \sum_{kk'\sigma\sigma'}
\delta_{\sigma\sigma'}{\!\!\!}
\int \frac{d\omega_{12}}{(2\pi)^2}
\bar{V}_{pk}  
\bar{G}_{p'k'pk}^{-+}(\omega_1) 
\bar{G}_{dp'k'}^{+-}(\omega_2), \\
\label{sh4}
\end{eqnarray}
%%%%%%%%%%%%%%%%%%%%%%%%%%%%%%%%%%%%%%%%%%%%%%%%%%%%%%%%
where $d\omega_{12}=d\omega_1 d\omega_2$. 

Now we examine the $k$-sum in eq.(\ref{sh4}). 
Appling eq.(\ref{g1}) for the summation over $k'$, 
essential part of calculations are reduced to a form 
%%%%%%%%%%%%%%%%%%%%%%%%%%%%%%%%%%%%%%%%%%%%%%%%%%%%%%%
\begin{eqnarray}
\nonumber
& &
\sum_{k'} V_{pk}  
G_{p'k'pk}^{-+}(\omega_1) G_{dp'k'}^{+-}(\omega_2) \\
& &
\rightarrow
\sum_{k'} |V_{p'k'}|^2 
g^{0\alpha\beta}_{p'k'p'k'}(\omega_1)
g^{0\gamma\delta}_{p'k'p'k'}(\omega_2). 
\label{shk}
\end{eqnarray}
%%%%%%%%%%%%%%%%%%%%%%%%%%%%%%%%%%%%%%%%%%%%%%%%%%%%%%%
Notice that 
the analytic properties of $g^{0\alpha\beta}_{p'k'p'k'}$ 
are determined by $\pm i\epsilon$ because of 
the free electrons in leads. 
Summation of the product of two $g^0$ 
with different frequencies $\omega_1$ and $\omega_2$ 
over $k'$ 
gives rise to a $\delta$-function singularity 
when $\omega_1$ is close to $\omega_2$,  
and a principal integration. 
This point is quite different from the one 
in $S_1$. 

In contrast, the summation over $k$ in eq.(\ref{sh4}) 
has the same character as the one 
previously discussed in $S_1$. 
Thus it does not produce the $\delta$-function singularity. 

To determine $S_{h1}$ 
we evaluate $2{\;}{\rm Re}(K_{1LL}+K_{1RR})$ 
%%%%%%%%%%%%%%%%%%%%%%%%%%%%%%%%%%%%%%%%%%%%%%%%%%%%%%%
\begin{eqnarray}
\nonumber
& &2{\;}{\rm Re}(K_{1LL}+K_{1RR}) = \\
\nonumber
& &2
\int \frac{d\omega}{2\pi} 
\left\{
\bar{T}(\omega) (1-\bar{T}(\omega)) (f_L-f_R)^2 
+ a(\omega) -b(\omega) \right\}. \\
\label{sh5}
\end{eqnarray}

%%%%%%%%%%%%%%%%%%%%%%%%%%%%%%%%%%%%%%%%%%%%%%%%%%%%%%%%
The term of $\bar{T}(1-\bar{T})$ and the $a(\omega)$-term 
originate form the $\delta$-function term where
%%%%%%%%%%%%%%%%%%%%%%%%%%%%%%%%%%%%%%%%%%%%%%%%%%%%%%%
\begin{eqnarray}
a(\omega) 
= \pi^2 \bar{\rho}(\omega)^2 (\bar{\Gamma}_L-\bar{\Gamma}_R)
\frac{\bar{\Gamma}_L\bar{\Gamma}_R}
{\bar{\Gamma}_L+\bar{\Gamma}_R}
(f_L-f_R). 
\label{sh6}
\end{eqnarray}
%%%%%%%%%%%%%%%%%%%%%%%%%%%%%%%%%%%%%%%%%%%%%%%%%%%%%%%%
The $b(\omega)$-term originates from the principal 
integration. 
We can prove that $b(\omega)$ 
completely agrees with $a(\omega)$: 
$b(\omega)=a(\omega)$. As a consequence 
$2{\;}{\rm Re}(K_{1LL}+K_{1RR})$ is characterized by 
$\bar{T} (1-\bar{T})$, as expected for shot noise. 

Moreover the same type calculations lead to that 
$2{\;}{\rm Re}(K_{1LR}+K_{1RL}) 
= -2{\;}{\rm Re}(K_{1LL}+K_{1RR})$. 
$S_{h1}$ becomes 
%%%%%%%%%%%%%%%%%%%%%%%%%%%%%%%%%%%%%%%%%%%%%%%%%%%%%%%
\begin{eqnarray}
S_{h1} =\frac{4e^2}{h}
\int d\omega 
\bar{T}(\omega) (1-\bar{T}(\omega)) (f_L-f_R)^2. 
\label{shb}
\end{eqnarray}
%%%%%%%%%%%%%%%%%%%%%%%%%%%%%%%%%%%%%%%%%%%%%%%%%%%%%%%%
We conclude that $S_{h1}$ has 
a shot-noise form, but 
characterized by the full $\bar{T}$. 
Employing eq.(\ref{tex}), 
the leading-order of $S_{h1}$ can be 
calculated at zero temperature as 
%%%%%%%%%%%%%%%%%%%%%%%%%%%%%%%%%%%%%%%%%%%%%%%%%%%%%%%
\begin{eqnarray}
S_{h1} =
\frac{4e^3}{h} |V| 
\left\{
\frac{1}{12} \left( \frac{eV}{\bar{\Gamma}} \right)^2
+ \frac{5}{12}(R-1)^2 \left( \frac{eV}{\bar{\Gamma}} \right)^2
\right\},
\label{ash1}
\end{eqnarray}
%%%%%%%%%%%%%%%%%%%%%%%%%%%%%%%%%%%%%%%%%%%%%%%%%%%%%%%%
where $\bar{U}/\pi\bar{\Gamma}=R-1$. 

Having addressed $S_{h1}$ for the bubble-diagrams, 
let us turn to $S_{h2}$ for the vertex-diagrams. 
We can confirm that for $S_{h2}$ only 
the term (a) in Fig.{\ref{diagram}} 
remains  
in the same way as $S_{2}$. 
Considering this point,
the second-order correction to $K_{2pp'}$ 
are evaluated. 
The obtained $K_{2pp'}$ for $p,p'=L,R$ 
decide $S_{h2}$ as follows, 
%%%%%%%%%%%%%%%%%%%%%%%%%%%%%%%%%%%%%%%%%%%%%%%%%%%%%%%
\begin{eqnarray}
S_{h2} = \frac{e^2}{\hbar} 
\int d\tau_1 d\tau_2 {\;} {\rm Re}F_h. 
\label{sh7}
\end{eqnarray}
%%%%%%%%%%%%%%%%%%%%%%%%%%%%%%%%%%%%%%%%%%%%%%%%%%%%%%%
$F_h$ is given by 
%%%%%%%%%%%%%%%%%%%%%%%%%%%%%%%%%%%%%%%%%%%%%%%%%%%%%%%
\begin{eqnarray}
\nonumber
F_h=
-i 4 \sum_{p'} \sigma_z^{p'p'} 
\left[ \sum_{k'} 
\bar{g}_{dp'k'}(\tau_1-0^{-})  
\bar{g}_{p'k'd}(0^{-}-\tau_2) \right] & \\
\nonumber
\times \bar{g}_d(0^{+}-\tau_1) 
\sum_{p} \sigma_z^{pp}
\sum_{k} \bar{V}_{pk}  \bar{g}_{dpk}(\tau_2-0^{+}) & \\
\nonumber
{\,}\times \bar{U}\bar{g}_d(\tau_1-\tau_2) 
\bar{U}\bar{g}_d(\tau_2-\tau_1). & \\
\label{sh8}
\end{eqnarray}
%%%%%%%%%%%%%%%%%%%%%%%%%%%%%%%%%%%%%%%%%%%%%%%%%%%%%%%
Following the discussion of the $k$-sum in $S_{h1}$, 
we find that 
$[\sum_{k'} ...]$ in eq.(\ref{sh8}) gives 
the $\delta$-function part and 
the principal-integration part. 

Taking this point into account, we perform calculations of
all Keldysh components of $F_h^{\alpha\beta}$, 
defined by $\tau_1=t_1^{\alpha}$ 
and $\tau_2=t_2^{\beta}$ 
in eq.(\ref{sh8}). 
We discuss 
relevant 
$\sum_{\alpha\beta}{\rm Re}F_h^{\alpha\beta}$ 
for $S_{h2}$. 
We find that 
the principal-integration parts
in $\sum_{\alpha\beta}{\rm Re}F_h^{\alpha\beta}$ vanish. 
Concerning the $\delta$-function part, 
${\rm Re}F_h^{--}$, ${\rm Re}F_h^{+-}$ and 
${\rm Re}F_h^{++}$ do not contribute 
to the leading-order of $S_{h2}$ in $V$. 
${\rm Re}F_h^{-+}$ determines 
the leading-order of $S_{h2}$
%%%%%%%%%%%%%%%%%%%%%%%%%%%%%%%%%%%%%%%%%%%%%%%%%%%%%%%
\begin{eqnarray}
\nonumber
S_{h2}=
-\frac{4e^2}{\hbar} c' 
\int d\omega_{1}d\omega_{2} 
( f_L(\omega_1 +\omega_2)-f_R(\omega_1 +\omega_2) )& \\
\nonumber
{\,}\times \bar{g}_d^{+-}(\omega_1) ( f_L(\omega_1)-f_R(\omega_1) ) 
\bar{g}^r(\omega_1) 
\times
\Gamma(\omega_2), &\\
\label{sh9}
\end{eqnarray}
%%%%%%%%%%%%%%%%%%%%%%%%%%%%%%%%%%%%%%%%%%%%%%%%%%%%%%%
where $c'=\pi\bar{\Gamma}^2\bar{U}^2/(2\pi)^4$.
$\Gamma(\omega)$ is the vertex function defined 
in eq.(\ref{np10}). 
Employing the asymptotic form of $\Gamma(\omega)$, 
$S_{h2}$ is evaluated as 
%%%%%%%%%%%%%%%%%%%%%%%%%%%%%%%%%%%%%%%%%%%%%%%%%%%%%%%
\begin{eqnarray}
S_{h2}=
\frac{4e^3}{h} |V| 
\left\{
\frac{1}{3} \left( R-1 \right)^2
\left( \frac{\bar{eV}}{\bar{\Gamma}} \right)^2 
\right\}+\cdots, 
\label{ash2}
\end{eqnarray}
%%%%%%%%%%%%%%%%%%%%%%%%%%%%%%%%%%%%%%%%%%%%%%%%%%%%%%%
where $\bar{U}/\pi\bar{\Gamma}=R-1$. 

Therefore, the asymptotic form of 
$S_h=S_{h1}+S_{h2}$ is determined as 
%%%%%%%%%%%%%%%%%%%%%%%%%%%%%%%%%%%%%%%%%%%%%%%%%%%%%%%
\begin{eqnarray}
\nonumber
& &
{\!\!\!\!\!\!\!\!\!\!\!\!\!\!\!\!\!\!}S_{h} =
\frac{4e^3}{h} |V| 
\left\{
\frac{1}{12} \left( \frac{eV}{\bar{\Gamma}} \right)^2 \right.
{\!\!\!\!}
+ \frac{5}{12}(R-1)^2 
\left( \frac{eV}{\bar{\Gamma}} \right)^2  \\
%
%\nonumber
& & {\qquad\qquad\qquad\qquad\ } 
{\!\!\!\!}
+\left. {\,}\frac{1}{3}{\,\,}(R-1)^2 
\left( \frac{eV}{\bar{\Gamma}} \right)^2
\right\}.  
\label{ash}
\end{eqnarray}
%%%%%%%%%%%%%%%%%%%%%%%%%%%%%%%%%%%%%%%%%%%%%%%%%%%%%%%%

Finally we summarize the results 
of the noise power $S=S_1+S_2$ 
and the shot noise $S_h=S_{h1}+S_{h2}$. 
Subscripts of $1$ and $2$ represent 
the contributions from the bubble-diagrams 
and the vertex-diagrams. 

As a check we begin with $S=S_1+S_2$. 
Taking the s-d limit leads to 
$\Gamma \rightarrow T_K$ and the Wilson 
ratio $R \rightarrow 2$. 
We find that 
the resulting asymptotic form of $S=S_1+S_2$ 
precisely agrees with the ones obtained 
as shot noise 
by using the fixed-point Hamiltonian
\cite{Sela06,Golub06,Gogolin06}. 
Up to ${\cal{O}}(V^3)$,
the second-order RPT 
indeed gives the correct result. 

We turn to $S_h$ proposed 
as the new formula of shot noise. 
Up to the leading order in $V$, 
$S_1=S_{1h}$ and $S_2=S_{2h}$ are clearly satisfied. 
Therefore 
for the symmetric case at zero temperature, 
$S_h$ precisely agrees with 
$S$ up to ${\cal{O}}(V^3)$. 
%%%%%%%%%%%%%%%%%%%%%%%%%%%%%%%%%%%%%%%%%%%%%%%%%%%%%%%
%%\begin{eqnarray}
%%\nonumber
%%S=&S_1&{\!\!\!\!}+S_2 \\ 
%%\nonumber
%%  &S_1&=\frac{4e^2}{h} \int d\omega \bar{T} (\omega) 
%%        (f_L (1-f_L)+f_R(1-f_R)) \\
%%\nonumber
%%  &   &+{\;}
%%        \frac{4e^2}{h} \int d\omega \bar{T} (\omega) 
%%        (1-\bar{T} (\omega))(f_L-f_R)^2 \\ 
%%\nonumber
%%  &   &=\frac{4e^3}{h} |V| \left\{
%%        \frac{1}{12} \left( \frac{eV}{\bar{\Gamma}} 
%%        \right)^2 
%%        {\!\!\!\!}+
%%        \frac{5}{12}(R-1)^2 \left( \frac{eV}{\bar{\Gamma}} 
%%        \right)^2 \right\} \\
%%\label{sums1} \\
%%%\nonumber
%%  &S_2&=\frac{4e^3}{h} |V| \left\{
%%        \frac{1}{3} \left( R-1 \right)^2
%%        \left( \frac{\bar{eV}}{\bar{\Gamma}} \right)^2 
%%        \right\} 
%%\label{sums2} \\
%%%
%%\nonumber
%%S_h{\!}=&S_{h1}&{\!\!\!\!}+S_{h2} \\ 
%%\nonumber
%%  &S_{h1}&=\frac{4e^2}{h} \int d\omega \bar{T} (\omega) 
%%            (1-\bar{T} (\omega))(f_L-f_R)^2 \\ 
%%\nonumber
%%  &      &=\frac{4e^3}{h} |V| \left\{
%%        \frac{1}{12} \left( \frac{eV}{\bar{\Gamma}} 
%%        \right)^2 
%%        {\!\!\!\!}+
%%        \frac{5}{12}(R-1)^2 \left( \frac{eV}{\bar{\Gamma}} 
%%        \right)^2 \right\} \\
%%\label{sumsh1} \\
%%%\nonumber
%%  &S_{h2}&=\frac{4e^3}{h} |V| \left\{
%%        \frac{1}{3} \left( R-1 \right)^2
%%        \left( \frac{\bar{eV}}{\bar{\Gamma}} \right)^2 
%%        \right\} 
%%%
%%\label{sumsh2}
%%\end{eqnarray}
%%%%%%%%%%%%%%%%%%%%%%%%%%%%%%%%%%%%%%%%%%%%%%%%%%%%%%%%
%
%
\subsection{Fano factor $F_b$}

In this section, we would like to argue the Fano factor
%%%%%%%%%%%%%%%%%%%%%%%%%%%%%%%%%%%%%%%%%%%%%%%%%%%%%%%
\begin{eqnarray}
F_b = \frac{S_h}{2eI_b}.
\label{fdf}
\end{eqnarray}
%%%%%%%%%%%%%%%%%%%%%%%%%%%%%%%%%%%%%%%%%%%%%%%%%%%%%%%%
%
Conventionally, noise power $S$ at zero temperature 
is treated as shot noise. 
However, we have proposed $S_h$ as shot noise 
at any temperature 
based on the nonequilibrium Kubo formula. 
%%As discussed previously, our idea has been 
%%justified 
%%in case of the symmetric case at zero temperature 
%%up to ${\cal{O}}(V^3)$. 
Therefore, it is natural to define 
the Fano factor with $S_h$ in eq.(\ref{fdf}). 
Here $I_b$ expresses 
the backscattering current
\cite{Sela06,Golub06,Gogolin06,Zarchin08} 
%
%%%%%%%%%%%%%%%%%%%%%%%%%%%%%%%%%%%%%%%%%%%%%%%%%%%%%%%
\begin{eqnarray}
I_b=\frac{2e^2}{h}V -I.
\label{f1}
\end{eqnarray}
%%%%%%%%%%%%%%%%%%%%%%%%%%%%%%%%%%%%%%%%%%%%%%%%%%%%%%%%
%
The $I$ in eq.(\ref{current2}) 
gives an expression of $2eI_b$ 
%%%%%%%%%%%%%%%%%%%%%%%%%%%%%%%%%%%%%%%%%%%%%%%%%%%%%%%

\begin{eqnarray}
2eI_b =
\frac{4e^3}{h} V
\left\{
\frac{1}{12} \left( \frac{eV}{\bar{\Gamma}} \right)^2
{\!\!\!\!}
+ \frac{5}{12}(R-1)^2 \left( \frac{eV}{\bar{\Gamma}} \right)^2
\right\}. 
%\nonumber
\label{f2}
\end{eqnarray}
%%%%%%%%%%%%%%%%%%%%%%%%%%%%%%%%%%%%%%%%%%%%%%%%%%%%%%%%
The backscattering current $2eI_b$ just 
corresponds to $S_{h1}$ for the bubble-diagrams in 
eq.(\ref{ash1}): $2eI_b =S_{h1}$ for $V>0$. 
Consequently $F_b =(S_{h1}+S_{h2})/2eI_b=1+S_{h2}/2eI_b$. 
Using expressions of $S_{h2}$ for 
the vertex diagrams in eq.(\ref{ash2}) and $2eI_b$ 
in eq.(\ref{f2}), we obtain $F_b$ 
%%%%%%%%%%%%%%%%%%%%%%%%%%%%%%%%%%%%%%%%%%%%%%%%%%%%%%%
\begin{eqnarray}
F_b =1+\frac{4(R-1)^2}{1+5(R-1)^2}. 
\label{f3}
\end{eqnarray}
%%%%%%%%%%%%%%%%%%%%%%%%%%%%%%%%%%%%%%%%%%%%%%%%%%%%%%%%

The first term in $F_b$ originates from $S_{h1}$. 
$S_{h1}$ is characterized by only $\bar{T}$. 
If correlation effect gives only a change  
from the bare $T^0$ to the full $\bar{T}$, 
$F_b=1$ would hold within ${\cal{O}}(V^3)$. 
Actually, $S_{h2}$ from the vertex-diagrams 
contributes to $F_b$. 
As a result it gives an enhancement factor of 
the second term in eq.(\ref{f3}). 
The vertex-diagram is known to describe 
a kind of the back-flow effect. 
The back-flow effect enhances 
the Fano factor for backscattering current. 
The form of the enhancement in $F_b$ 
has a universal feature that 
only the Wilson ratio determines it for any $U$. 

Here we check the resulting $F_b$  
in the limit of 
$U \rightarrow \infty$ and $U \rightarrow 0$. 
Appling limiting values of the Wilson ratio $R$ 
for $F_b$ leads to 
%%%%%%%%%%%%%%%%%%%%%%%%%%%%%%%%%%%%%%%%%%%%%%%%%%%%%%%
\begin{eqnarray}
F_b= \left \{
\begin{array}{clll}
{\ }\displaystyle{\frac{5}{3}} &{\ \ }& U \rightarrow \infty & R=2 \\
 &{\ \ }& & \\
{\ }1           &{\ \ }& U \rightarrow 0      & R=1 \\
\end{array}
\right. 
\label{f4}
\end{eqnarray}
%%%%%%%%%%%%%%%%%%%%%%%%%%%%%%%%%%%%%%%%%%%%%%%%%%%%%%%%
$F_b$ indeed reproduces 
the universal fractional value of $F_b=5/3$ 
derived in the s-d limit\cite{Sela06,Golub06,Gogolin06}, 
and a naively expected value $F_b=1$ 
for a noninteracting system. 
Therefore $F_b$ given by eq.(\ref{f3}) 
is an extension for any $U$ 
from the already obtained value of $F_b=5/3$ 
in the s-d limit. 

Before closing discussion, we touch on 
an application of $F_b$. 
By using $F_b$, 
we may determine 
the Wilson ratio directly from experiments as follows, 
%%%%%%%%%%%%%%%%%%%%%%%%%%%%%%%%%%%%%%%%%%%%%%%%%%%%%%%
\begin{eqnarray}
R=1+\sqrt{\frac{F_b -1}{9-5F_b}}.
\nonumber
\label{f5}
\end{eqnarray}
%%%%%%%%%%%%%%%%%%%%%%%%%%%%%%%%%%%%%%%%%%%%%%%%%%%%%%%
%
%
%
%
%
\section{Summary}

The nonequilibrium Kubo formula $S_h=S-4k_{\rm B}TG$ 
allows us to propose $S_h$ as the new definition 
of shot noise in general. 
Experimentally measured $S-4k_{\rm B}TG$ 
can be compared with 
a theoretical prediction of $S_h$ at any temperature. 
Therefore, the nonequilibrium Kubo formula 
thus opens a new approach to 
studies of shot noise in correlated systems 
at any temperature and any bias voltage. 

Then, using this approach 
shot noise through a quantum dot 
in the Kondo regime has been investigated. 
For simplicity, the symmetric case has 
been discussed at zero temperature. 
At $T=0$, the nonequilibrium Kubo formula gives $S_h=S$. 
We have thus analyzed both $S$ and $S_h$, 
up to ${\cal{O}}(V^3)$ which is 
the leading order in $V$ for the symmetric case. 
The renormalized perturbation theory (RPT) 
has enabled us to obtain the exact asymptotic 
form of $S$ and $S_h$. 
Both of $S$ and $S_h$ are expressed by 
two-particle Green functions. 
Thus they have two types of 
contributions from the bubble-diagrams
and the vertex-diagrams 
as $S=S_1+S_2$ and $S_h=S_{h1}+S_{h2}$. 
It has been shown $S_{1}=S_{h1}$ and 
$S_{2}=S_{h2}$ up to ${\cal{O}}(V^3)$. 
We have concluded that 
$S_h$ indeed equals $S$ at $T=0$ 
which was conventionally defined as shot noise. 
Finally we have pointed out that 
$S_{h2}(=S_2)$ from the vertex-diagrams 
leads to the universal enhancement factor 
in the Fano factor: 
$F_b=1+4(R-1)^2/(1+5(R-1)^2)$. 
This expression includes 
the result of $F_b=5/3$ in the Kondo limit 
by using the Wilson ratio $R=2$. 

Furthermore, 
we have found that $S_1$ splits 
into the thermal-noise part and 
the shot-noise part with the full 
transmission probability $T$. 
Here, let us recall the fact of 
$S=S_t(T)+S_s(T)+\Delta S$. 
In fact, $S_1=S_t(T)+S_s(T)$. 
Thus, $\Delta S$ describes 
nothing but the contribution 
from the vertex-diagrams: $S_2=\Delta S$. 
As discussed previously, concerning $\Delta S$ 
it is not clear whether noise power $S$
at a general temperature is split into 
the thermal-noise part and the shot-noise part. 
However, it is impossible to ignore the effect of 
$\Delta S$, even at zero temperature 
because $S_{h2}=S_2=\Delta S$ is 
always relevant as in the universal enhancement 
factor in $F_b$. 
\section{ACKNOWLEDGEMENT}

The author would like to thank K. Ueda, Y. Tokura 
, A. Oguri and Y. Utsumi for helpful discussions. 
%%%%%%%%%%%%%%%%%%%%%%%%%%%%%%%%%%%%%%%%%%%%%%%%%%%%%%%%%%%%
%%%%%%%%%%%%%%%%%%%%%%%%%%%%%%%%%%%%%%%%%%%%%%%%%%%%%%%%%%%%

\end{document}